\documentclass[article]{rstransa}%%%%where rstransa is the template name

\usepackage{color}            % to get colors
\definecolor{red}{rgb}{1,0,0} % defines 'red'
\definecolor{blu}{rgb}{0,0,1} % defines 'blue'
\definecolor{bla}{rgb}{0,0,0} % defines 'black'
\definecolor{pur}{rgb}{0.5,0,0.6} % defines 'purple'
\definecolor{gre}{rgb}{0,0.6,0} % defines 'green'

\usepackage{lipsum}
\usepackage{multirow}
\usepackage{appendix}

\titlehead{Research}

\begin{document}

%%%% Article title to be placed here
\title{Investigating the formation of small Solar System objects using stellar occultations by satellites: present, future and its use to update satellite orbits }

\small\author{
F.~Braga-Ribas$^{1,2,3}$, %wrote the paper, calculated the positions, etc.
F.~Vachier$^{4}$, %0000-0002-4289-4466  GENOID
J.~Desmars$^{4,5}$, % NIMA ephemeris 
G. Margoti$^{1,2,3}$,
% calculated the satellite positions
B.~Sicardy$^{4}$ % reviewd the paper, group leader
} 
%%%%%%%%% Insert author address here
\address{\small{
$^{1}$Federal University of Technology - Paraná (PPGFA/UTFPR), Curitiba, PR, Brazil;
$^{2}$Laboratório Interinstitucional de e-Astronomia - LIneA, Rio de Janeiro, RJ, Brazil;
$^{3}$Observatório Nacional/MCTI, Rio de Janeiro, RJ, Brazil;
$^{4}$Institut de Mécanique Céleste et de Calcul des Éphémérides, IMCCE, Observatoire de Paris, PSL Research University, CNRS, Sorbonne Universités, UPMC Univ Paris 06, Univ. Lille, France;
$^{5}$ Institut Polytechnique des Sciences Avancées IPSA, 94200 Ivry-sur-Seine, France;
}}
%%%% Subject entries to be placed here %%%%
\subject{planetary astrophysics, orbits, minor planets.}

%%%% Keyword entries to be placed here %%%%

%%%% Insert corresponding author and its email address}
\corres{Felipe Braga Ribas\\
\email{fribas@utfpr.edu.br}}

%%%% Abstract text to be placed here %%%%%%%%%%%%
\begin{abstract}
% 200 words
The history of the outer solar system is intrinsically related to the  Giant Planets migration.  A massive disk of material  within a radius  of 30~au was scattered during the planetary migration, creating different dynamic populations in the Transneptunian region. They were formed in a collisional environment when massive collisions allowed them to grow and form  much smaller moons  than the primary body. The dynamical group, known as the Cold Classicals, was formed in a sparse disk from 42 to about 47~au and did not suffer much from planet migration. Observations show that many of Cold Classical are binary, consistent with the streaming instability process. The stellar occultation technique, with a spatial resolution of a few kilometres, can be used to search for binaries where other techniques are unable to do so, and to characterise the known satellites  of Trans-Neptunian Objects (TNO), constraining their formation scenarios. We review here the first stellar occultations by TNO's satellites (besides Charon), discuss the methods used to detect these events. We also fit new orbital elements and system mass for Vanth (Orcus/1) and Weywot (Quaoar/1), finding reasonable solutions for pure Keplerian orbits. Finally, we discuss the prospects regarding the stellar occultations by TNO binaries and their implications for the study of the history of the Solar System.

\end{abstract}
%%%%%%%%%%%%%%%%%%%%%%%%%%%

%%%%%%%%%% Insert the texts which can accommodate on firstpage in the tag "fmtext" %%%%%
\maketitle
\section{Introduction}
\label{sec:Introduction}

The complete history of the Outer Solar System's current structure is yet under discussion. In short, many works have shown that the Giant Planets were formed in a tighter configuration closer to the Sun.  
After a few hundred million years, they migrated outwards, scattering a massive disk of planetesimals that existed inside 30~au \cite{Levison03,Gomes04,Tsiganis05,Nesvorny16}. %Nice model
These objects fed the Kuiper Belt's dynamical regions occupied by the Hot Classicals (HC), Resonant populations, such as the Plutinos, and the Scattered Disk Objects (SDO) \cite{Gladman08,Gladman21}.
Another dynamical group, the Cold Classicals (CC), had a different history  \cite{ParkerKavelaars10,ThirouinSheppard2019}. They were formed in a sparse disk from 42 to about 47~au and did not suffer much from planet migration.

% The paragraph was transported here to reply to item A of the second reviewer.
Stellar occultation has proven to be a powerful technique for studying the Transneptunian region \cite{Sicardy2024,Ortiz20}. With a spatial resolution of a few kilometres, independent of the object's distance,  which is ultimately limited by the Fresnel diffraction and stellar diameter,  it has led to the discovery of rings around small objects, such as the Centaur (10199) Chariklo \cite{Braga-Ribas2013}, the dwarf-planet (136108) Haumea \cite{Ortiz2017} and the TNO (50000) Quaoar \cite{Morgado2023, Pereira2023}. It has also been used to monitor Pluto's atmosphere \cite{Meza19}, study TNO's topography \cite{DiasOliveira17, Rommel2023}, and provide precise astrometric positions \cite{Rommel2020}. As discussed further in this Section, this resolution can be used to study the close environment of small objects of the outer Solar System and their satellites. It also allows us to search for putative companions where other techniques are unable to investigate. This comes in a context where the number of TNO satellites became a fundamental information to constrain the formation scenarios.

The Cold Classicals are small bodies with low orbital inclinations and eccentricity,
formed in a low collisional environment. Observations show that more than 30$\%$ of Cold Classicals are found to be binaries, with equal size and similar colour of the components  \cite{Fraser17}.
All this is consistent with an in situ formation through the streaming instability process, which predicts that most (if not all) planetesimals are formed in binary or multiple systems \cite{NesvornyVok19}. 

In contrast, the Hot Population objects (HP)\footnote{By Hot Population we refer to the Hot Classicals, Resonant population, and the Scattered Disk Objects.} were formed and grew in a collisional environment (a massive planetesimal disk) and then transported to their current positions through scattering encounters with Neptune \cite{Levison08}. Differently from the CCs, the majority of the moons of the %Hot population 
HP are much smaller than the primary (with a radius ratio <~0.5), and almost all of the most prominent members (with radii R~>~500 km) have known satellites. This indicates that they were accreted from a disk around the primary, which, in turn, was formed from a collision of two planetesimals \cite{Leinhardt10,ThirouinSheppard2018}, a condition only found in the massive planetesimal disk, prior to their implementation at their current location. It was shown that 50\% of the bound binaries with $a_B/R_B$~<~30 (where $a_B$ is the binary semi-major axis and $R_B=(R{_1^3}+R{_2^3})^{1/3}$, being $R_1$ and $R_2$ the radii of each component) survive the implementation to the HP region. The wide binaries instead did not survive, so they may have originated before scattering in a rarefied %Sparse, loose, void, rarefied, less dense
part of the disk beyond 30~au \cite{NesvornyVok19}.

Most of this scenario is based on the interpretation of the numerical simulations, which are, in turn, bounded by direct observations of the objects, e.g., their relative sizes, colours, and presence and orbit of satellites. In many cases, sizes are based on assumptions of colours and albedos or via observations on the thermal region of the spectrum, and most of the satellites or binaries were discovered using high-resolution images. These techniques have their caveats and limitations. For instance, observations have shown that at least 30\% of the CCs are binaries   \cite{Grundy19}, but this is based mainly on Hubble Space Telescope (HST) observations. HST's best resolution is 40~milli-arcsec (mas) in the plane of the sky, which translates to a spatial resolution of about 1200~km at 42~au. Therefore, binaries or satellites closer to this distance are blended in the images, so they can not be firmly detected with direct images \cite{Porter24}. 

Considering the context presented above, some unanswered questions that can help narrow the formation scenarios and determine the global properties of TNBs remain to be addressed:
\begin{itemize}
    \item Are all the Cold Classicals binaries?
    \item Are small satellites a rare property in the Cold Classicals region?
    \item Is there a gap between being a  (separated)  binary system and a  contact binary  object?
    \item Are the hot population moons similar to the primary in terms of composition and density?
    \item Do the satellites have the same albedo as the primary? Therefore, are albedo-based sizes a good estimation?
    \item How many satellites were formed after the implantation in the hot population region?
    \item Do rings exist in the Cold Classicals region, or do they need a collisional active environment to be formed?
\end{itemize}

The  stellar occultation technique can be used to probe the TNO's close environments,  study the TNO's satellites, search for unknown companions or  reveal  binarity. The chances of detecting an unknown component are small; still, detecting one of the components already brings  valuable information, such as the position of the secondary relative to the primary. The precision so obtained is comparable to that derived with high-resolution images from ground-based or space telescopes. 
It can also be related to information from other techniques, such as astrometric wobbling and thermal size determination \cite{Ortiz20}.

Observation of occultations by TNO satellites or binaries (TNOBs, for short) is still rare. A few stellar occultations by Charon, Pluto's main satellite, were used to determine its size and constrain its orbit \cite{Sicardy11}. The first stellar occultation observed by a TNOB, after Charon, was caused by Vanth, the satellite of (90482) Orcus, on March 1, 2014 \cite{BragaRibas17}. In March 2017, a  fortuitous  observation of another occultation by Vanth allowed determine its size, assuming spherical, with a diameter of 443~$\pm$~10~km \cite{Sickafoose19}. Two stellar occultations by Hi'iaka, the biggest satellite of (136108) Haumea, were predicted and detected in April 2021, constraining the satellite's size and shape \cite{Fernandez21} (Fernandez-Vallenzuela, 2024, \textit{submitted}). The Plutino (523764) 2014~WC$_{510}$ was discovered to be a similar-sized binary during a stellar occultation in December 2018 \cite{Leiva20}. Observations of occultations by the Hot Population satellites of (50000) Quaoar \cite{Fernandez23weywot}, (38628) Huya \cite{Rommel25}, (82075) 2000~YW$_{134}$ \cite{Vara23}, (19521) Chaos \cite{Leiva23}, and  (120347) Salacia  have been reported\footnote{From \cite{Braga-Ribas2019}, \url{https://occultations.ct.utfpr.edu.br/results/} updated on September 07, 2024.}.

In this section, we have discussed the connection between the Solar System's history and the formation of binaries, and we have provided the current list of stellar occultations by TNO satellites that we are aware of. In Section~\ref{sec_methods}, we will discuss the methods used to detect secondary events, either by chance or predicted from satellite ephemerides.
In Section~\ref{sec_Orbit}, we will present new orbit and mass parameters obtained for the Orcus and Quaoar systems based on positions derived from stellar occultations. The implications of these results will be addressed in Section~\ref{sec_discuss}, where we will also consider why we the number of recorded stellar occultations by TNOBs (especially for the CCs) will greatly increase in the near future.

\section{Observing Stellar Occultations by TNO Satellites}
\label{sec_methods}

Detecting a stellar occultation by a TNO satellite can result from i) a lucky observation or ii) a predicted event.

\subsection{Lucky detection}

When an occultation event is recorded, a secondary drop in the light curve may be detected. If the secondary event is as deep as the (expected) flux drop caused by the main body, it may be caused by a satellite. In the opposite case, if the drop is not as deep as expected by the primary body, then it may stem from either a semi-transparent ring or from the presence of a companion star. An unexpected secondary drop may be associated with a known satellite if the chord length and relative distance to the primary can be correlated to the known properties of the moon \cite{Rommel25}.

\begin{figure}
    \centering \includegraphics[width=0.9\hsize]{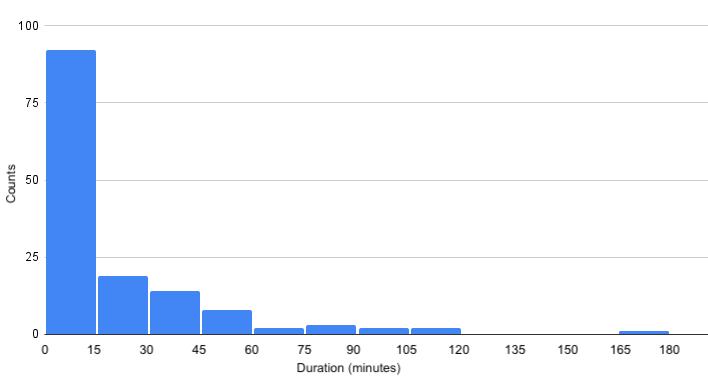}
    \caption{Histogram of the number of known TNOBs with known semi-major axis and the total observation interval, in minutes, around the main body event to reach the satellite distance, considering the typical event velocity of 20~km~s$^{-1}$.  One-hour long observation can probe 92\% of the known TNOBs orbits on both sides relative to the central body.}
    \label{fig_histminutes}
\end{figure}

To increase the chances of secondary detection, observations should last  long enough  to probe the object's environment. The duration will depend on the expected semi-major axis of the satellite and the event's velocity (typically 20 km s$^{-1}$). The  currently known TNO binary with the largest separation (more than 103500~km) is  2001~QW$_{322}$. This corresponds to almost three hours of observations for the star to cross both sides of the satellite's orbit, however, this is an extreme case. Out of the 145 known semi-major axes of TNOBs\footnote{Compilation made by Johnston, Wm. Robert, \url{https://www.johnstonsarchive.net/astro/asteroidmoonsall.html} updated on May 27, 2024.}, 65\% need only 15 minutes of observation, 73\% need  up to  30 minutes and  92\% less than an hour  (Fig. \ref{fig_histminutes}).

To estimate the largest distance of a putative satellite, we can use the object's half Hill's sphere, which depends on its distance to the Sun (or other nearby objects, such as a Giant Planet \cite{Araujo16}) and its mass. Considering their distances from the Sun, using this criterion implies an unrealistic long observation runs of hundreds of hours. Meanwhile, it is expected that a satellite will disintegrate if it gets too close to the main body (roughly 2.5 radii of the main body). This is the Roche limit, which depends on the relative masses and radius of the secondary \cite{Sicardy2020}. But this is just a crude estimation, as the object's cohesion force can be strong enough to avoid its disruption, especially for similar-sized bodies.

Another approach is to place a number of telescopes in a "fence" distribution perpendicular to the shadow motion, which is typically along the East-West direction. 
The telescopes must be spaced by at maximum half of the object's expected diameter, over a thousand kilometres, centred at the predicted shadow path. This is especially suitable for surveying the Cold Classicals searching for close binary objects \cite{Leiva20}.

\subsection{Prediction}

To predict a stellar occultation, both i) the position of the target object in time (i.e., its ephemeris) and ii) the position of a star on its path, for a given observer, need to be known with a precision equivalent to the apparent size of the object in the plane of the sky, roughly, tens of mas.  The \textit{Gaia} catalogue provides position, for most stars brighter than magnitude 21 \cite{GaiaColab2023},  with uncertainties that are much smaller than the apparent sizes of the TNOs. The challenge is thus to determine the satellite ephemeris. For that purpose, precise positions relative to the central body are needed. They are usually obtained with space telescopes (like HST) or large ground-based telescopes (like Keck) with high-resolution imaging capacity, i.e., adaptative optics.  This means a high pressure put on these instruments, and thus difficulties in obtaining observing time. 

In this work, we use the GENOID algorithm (GENetic Orbit IDentification)\cite{Vachier12,Vachier22} to fit an orbital model on the known satellite observations in the study. It is a genetics-based algorithm to find the most appropriate set of dynamic parameters, i.e., minimum $\chi^2$. We start modelling the satellite's motion around the main body with a Keplerian model \cite{Yang16}; if the residuals are significant with respect to the quality of the data (i. e. tens of mas), non-Keplerian perturbations such as the  dynamical oblateness $J_2$ of the primary and the satellite masses can be considered. 

GENOID can give precise ephemeris of the satellite relative to the main body, depending on the number and quality of the available positions,  through the Miriade ephemeride hosted by the IMCCE laboratory at Paris Observatory\footnote{\url{https://ssp.imcce.fr/forms/ephemeris}}.  Like all orbit-fitting algorithms, GENOID requires at least three sufficiently precise primary-to-satellite relative positions, well-distributed in time and space, to obtain an acceptable solution for orbital parameters along with their uncertainties. It is able to deliver robust results when dealing with small and poorly distributed datasets, investigating all the space of parameters. However, if the dataset is sufficient, it offers little advantage compared to less computationally demanding methods (e.g., variational methods \cite{Proudfoot24}).

\begin{figure}
    \centering     \includegraphics[width=0.45\hsize]{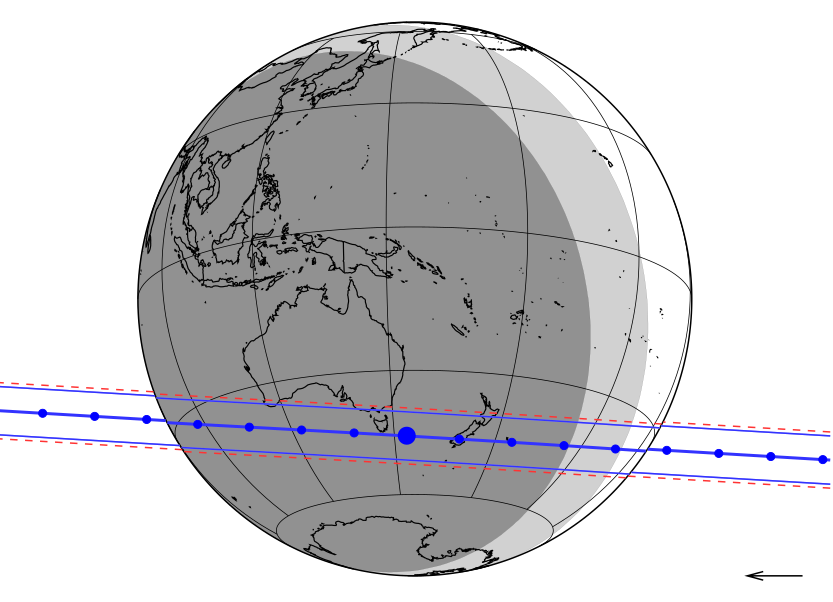}
    \includegraphics[width=0.45\hsize]{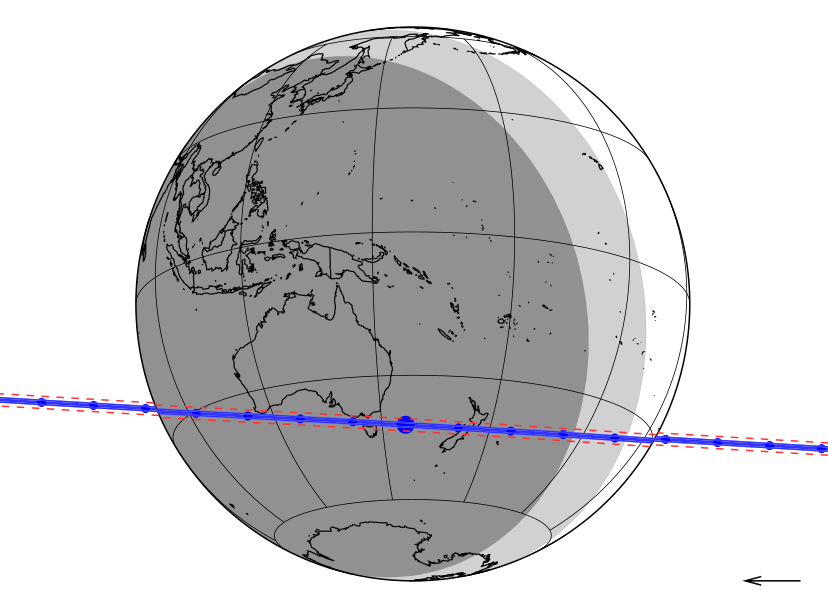}
    
    \caption{Prediction of an occultation by Quaoar (left) and its satellite Weywot (right) on May 26, 2023. The parallel blue lines, separated by the object's expected diameter, are the shadow paths on Earth. The arrow indicates the direction of  the shadow  motion, the dots are separated by one minute. The 1$\sigma$ expected uncertainty is shown as red dashed lines and are plotted relative to the shadow limits (not its centre).
    }
    \label{fig_mapsQuaoarWeywot}
\end{figure}

For the on-sky position calculation, we used the NIMA ephemeris \cite{desmars2015}, which is designed to provide precise ephemeris for a short period, using Minor Planet Center and proprietary data, as well as occultation-based astrometry when available. The relative offset of the satellite, calculated with GENOID, is applied to the main body's ephemeris, calculated with NINA, to obtain the satellite's position and then compare it to the \textit{Gaia} star catalogue. The predictions are available on the Lucky Star webpage\footnote{\url{https://lesia.obspm.fr/lucky-star/}}. This method was used to successfully predict the two occultations observed by Hi'iaka in April 2021\footnote{\url{http://fredvachier.free.fr/binaries/occult/2021/2021-04-06T23_33_52_136108_NIMAv72020Oct21_Binary_solution_1_Kepler/}}$^,$\footnote{\url{http://fredvachier.free.fr/binaries/occult/2021/2021-04-16T06_29_54_136108_NIMAv72020Oct21_Binary_solution_1_Kepler/}} and  occultations by Vanth and Weywot, wich will be detailed in this work. Figure~\ref{fig_mapsQuaoarWeywot} gives an example of a predicted occultation involving Quaoar and its satellite Weywot, which eventually yielded the detection of both objects on May 26, 2023. 

 \section{Orbits}
\label{sec_Orbit}

As previously mentioned, the satellites Vanth (Orcus/1) and Weywot (Quaoar/1) have already been detected  during  stellar occultations. Here, we use the relative positions derived from these detections to derive new orbits. 
The relative position can be accurate to the sub-mas level when multiple chords are detected  for both the main body and the satellite, that is, when both objects are detected from several sites during the same event. The on-sky positions of their respective centres can then be pinned down with an accuracy of a few kilometres.

If they are detected from only one site, then two solutions are possible, one with the centre to the North of the chord and one with the centre to the South of the chord \cite{Rommel2020}.  In this case, due to the unknown   size of the satellite, the centre uncertainty may be bounded by the expected object's radius (assumed to be spherical), corresponding to a few mas at the TNOs distances. However, both Vanth and Weywot may be non-spherical, which can produces an elongated limb of apparent semi-axes $a'$ and $b'$ and introduces an additional cause of error. We will have a smaller error in the direction of the apparent motion (along-track) and a larger error in the perpendicular direction of motion (across-track), see \cite{Ferreira20}. In an extreme case, where the limb apparent oblateness is $\epsilon' = (a'-b'/a') = 0.5$, this will induce an error of about half of $a'$, i.e. an additional error of at most  $R_{\rm equiv}/(2\sqrt{1-\epsilon}) = R_{\rm equiv}/\sqrt{2}$ for $\epsilon < 0.5$.

When only the satellite is detected occulting the star, the relative position is obtained from the ephemeris of the main body. (i) If the main body ephemeris uses positions derived from occultations \cite{Camargo18}, the relative positions may be obtained, depending on the number of positions and their time-span, with uncertainties in the order of the mas, i.e. a few tens of kilometres. (ii) On the other hand, if only classical astrometry  of the system  is available, then the primary body position can have an uncertainty of tens of mas, i.e. hundreds of kilometres. In this case, the position of the main body also needs to be corrected from the photo-centric offset caused by the satellite contribution to the total flux \cite{Rossi14},  as the ephemeris provides not the position of the central body but for the system photocenter. Therefore, the relative uncertainty may be poorly constrained. 

Using any relative position is important as they can constrain the orbital solution when only a few relative points are available, which is usually the case.  Thus, the mere fact that the satellite occulted the star at a given time already provides the satellite position with an uncertainty comparable to its diameter, corresponding to a few mas on the sky, as mentioned earlier. This is comparable to the accuracy obtained with space- or ground-based telescopes, which is thus useful in improving the satellite's orbital solution. Further analysis of the detected event can reduce the uncertainty; nevertheless, this extra precision will have little impact on the quality of the orbit fit, considering that it also uses relative positions derived from direct imaging with their own uncertainties.

 In this work, we use satellite offset positions 
$(X,Y)$\footnote{ The offset (X,Y) gives the apparent differential coordinates of the satellite relative to the primary, as seen in the sky plane.} 
relative to the  primary calculated from the reported occultations, limiting our precisions to the satellite's radius
  (assuming a spherical body, as discussed earlier), together with    published
offsets\footnote{ Here we used the offset values available in Grundy's web page (\url{https://www2.lowell.edu/users/grundy/tnbs/}), which were reanalysed with their tools to provide a consistent data set, as explained in \cite{Grundy19}.}, 
to fit the dynamical parameters of the system:  
mass ($M$), 
orbital period ($P$),
semi-major axis ($a$),
eccentricity ($e$),
inclination ($i$), 
longitude of the node ($\Omega$), 
argument of pericentre ($\omega$), 
and time of passage  at pericentre  ($t_{\rm p}$). If needed, the algorithm also fits the satellite masses and the $J_2$ factor related to the object's oblate gravity field \cite{Berdeu22}.
After setting a grid of test values, they are refined, generation after generation, until no significant improvement is obtained, i.e., the residuals of the data points compared to the fitted orbit cease to reduce.

\subsection{Vanth (Orcus/1)}
\label{subsec:Vanth}

Orcus is one of the largest Plutino objects, which is an object locked in the 2:3 mean motion resonance with Neptune. It has an estimated diameter of 910~$\pm$~50~km and a high albedo of 0.23~$\pm$~0.02 \cite{Brown23}. Its only known satellite, Vanth, is about 2.4 magnitudes fainter, with an albedo-dependent diameter from 280 to 380~km \cite{Brown10}. Previous works \cite{Brown10, Carry11, Grundy19} reported an almost circular orbit, with a period of 9.5 days and a semi-major axis of about 9000~km. 

Two stellar occultations by Vanth have been detected so far. The first one was also the first stellar occultation by a TNO satellite ever observed, apart from the Pluto-Charon system \cite{Sicardy11}, and was detected in March 2014. The second detection occurred in March 2017, when two chords allowed an estimate of its diameter, assuming a circular limb to be 443~$\pm$~10~km \cite{Sickafoose19}.

\subsubsection{March 01, 2014}

The first stellar occultation by Vanth was detected on March 1$^{\rm st}$, 2014, at the Nayoro Observatory, Hokkaido, Japan \cite{BragaRibas17}.
%\footnote{Nayoro Observatory, Faculty of Science, Hokkaido University, Nayoro, Hokkaido, Japan, Long.: 142$^\circ$ 28' 58" E, Lat.: 44$^\circ$ 22' 25.1" N, height: 161~m WGS84.}. 
Predictions made with the methods described in \cite{Camargo14} identified an occultation of the TYC 5476-00882-1 star (V$\rm{_{mag}}$ = 12.1) by Orcus crossing the South of Australia and New Zealand. Using GENOID-based ephemeris for Vanth, we found that the satellite could cross Japan. An extensive campaign was set to detect both objects, but eventually, only the satellite was detected.

From the observation, a $73.7~\pm~3.2~\rm km$ chord was obtained. Therefore, two on-sky positions for Vanth's centre are equally possible, one considering the centre to the North of the observed chord and the other where the centre is to the South of it. The astrometric position is derived by fitting the 443~km diameter from \cite{Sickafoose19} and considers the propagated Gaia DR3 position.
Using the latest NIMAv10 ephemeris for Orcus, with an estimated uncertainty of 4 mas, we can calculate the Orcus-to-Vanth relative position (Table \ref{tab_Vanthpos2014}). Orcus ephemeris is based on direct images of the system; thus, it represents the  photocentre  motion in the sky. To have the correct relative position, we used Equation A.13 from \cite{Rossi14}, considering the aforementioned expected relative position of 0.24~$\pm$~0.01~arcsec, the expected magnitudes 19.05~$\pm$~0.01 for Orcus and 21.4~$\pm$~0.1 for Vanth, to compute the relative fluxes. This resulted in a photocentre offset on the position of Orcus in the direction to Vanth of $0.023~\pm~0.003$~arcsec. The corrected relative position, accounting for the main body uncertainty, is presented in Appendix \ref{app_positions}, Table \ref{tab_posVanth}.

\begin{table}[ht]
    \centering
    \begin{tabular}{c c c c c}

    \multicolumn{5}{c}{Vanth: March 01, 2014 astrometry}  \\

    \hline \hline
    
    \multirow{2}{*}{Solution}   & 
    Date and Time (UT)  &
    Right ascension (h m s)    &  
     Offset  &
    Uncertainty \\ 
    
     &    
     Julian date &
     Declination  ($^\circ$ ’ ”)  &  
     $ (X,Y)$ (km) &
     (km)      \\
     \hline \hline

    \multirow{2}{*}{North}   &
    2014-03-01 16:19:01.120 &
    09 58 22.55266 $\pm$ 0.19 &
    2760.8 &
    6.3 \\

      &
     2456718.1798740742 &
    -08 16 55.1837 $\pm$ 0.116 &
    6530.1 &
    3.8 \\
     \hline 

    \multirow{2}{*}{South}   &
    2014-03-01 16:19:01.120 &
    09 58 22.55234 $\pm$ 0.20 &
    2600.5 &
    6.8 \\

      &
     2456718.1798740742 &
    -08 16 55.19536 $\pm$ 0.087  &
    6133.0 &
    2.8 \\
    \hline
    \end{tabular}
    \caption{Vanth's ICRS position for each possible solution, i.e., body centre to the North or South of the chord, for the event date, time and an object distance of 47.08 au. The propagated Gaia DR3 star position is RA: 09h 58m 22.54817s~$\pm$~0.0302 mas, DEC: -08$^\circ$ 16' 55''.34010~$\pm$~0.0315 mas.
     The uncertainties of Vanth's Right Ascension and Declination ($\Delta\alpha(\cos\delta),\Delta\delta)$) are given in mas. 
    The offset position $(X,Y)$ of Vanth is given  in the tangent plane aligned with the equatorial coordinates  with respect to the main body ephemeris (NIMAv10). Contrarily to Table \ref{tab_posVanth}, the 1$\sigma$ uncertainties on $(X,Y)$ are based here only on the expected satellite diameter, assumed to be spherical, fitted to the observed occultation chord.
    }
    \label{tab_Vanthpos2014}
\end{table}

\subsubsection{March 07, 2017}
%2017/03/07 description, results

The occultation by Vanth on March 07, 2017, involved a (previously unknown) binary star. The event was fortuitously detected from two sites,  and high-resolution images of the target star provided subsequently their  positions relative to the photocentre of the system, see details in \cite{Sickafoose19}. 
The authors used astrometric positions obtained the night of the occultation to calculate the Orcus-Vanth system position relative to the bright star and verified that the obtained value agrees with the detected occultation. With that, they derived an Orcus-Vanth relative position of $\Delta\alpha(\cos\delta) = -0.0445~\pm~0.010$~ arcsec and $\Delta\delta = -0.2362~\pm~0.001$~arcsec. We have also corrected the relative position by the photocentre offset in the Orcus-Vanth direction by $0.018~\pm~0.002$~arcsec to derive the  $(X,Y)$ positions  provided in Appendix \ref{app_positions}, Table \ref{tab_posVanth}.

Using the reported times and the stars' photocentre, we calculated the satellite's astrometric as well as the relative to Orcus position considering NIMAv10 ephemeris (Table~\ref{tab_Vanthpos17}).

\begin{table}[ht]
    \centering
    \begin{tabular}{c c c c c} 

    \multicolumn{5}{c}{Vanth: March 07, 2017 astrometry}  \\

    \hline \hline
    
    \multirow{2}{*}{Solution}   & 
    Date and Time (UT)  &
    Right ascension (h m s)    &  
     Offset &
    Uncertainty \\ 
    
     &    
     Julian date &
     Declination  ($^\circ$ ’ ”)  &  
     $ (X,Y)$ (km) &
     (km)      \\
     \hline \hline

    \multirow{2}{*}{Photocentre}   &
    2017-03-07 06:56:37.440 &
    10 08 18.547109 $\pm$ 0.76 &
     -1618.8 &
    1.7 \\

      &
     2457819.7893222221 &
    -09 40 14.010687 $\pm$ 0.46 &
     -6921.5 &
    2.7 \\
    \hline
    \end{tabular}
    \caption{Similar to Table~\ref{tab_Vanthpos2014}, but for the occultation of 2017. The satellite position considers the binary star's photocentre, RA = 10h 08m 18.54297 s $\pm$ 0.76 mas; DEC = -09$^{\circ}$ 40' 14''.14747 $\pm$ 0.46 mas (see text and \cite{Sickafoose19}). The  offset $(X,Y)$ is based on the  NIVAv10 ephemeris, i.e., the system photocentre. The object was at a geocentric  distance of 47.14 au.}
    \label{tab_Vanthpos17}
\end{table}

\subsubsection{Orbital elements}

Using GENOID, the aforementioned occultation-derived positions, and 14 relative positions from high-resolution images \cite{Brown10, Carry11, Brown23},  we obtained pure Keplerian orbits of Vanth around Orcus. Both mirrored, prograde and retrograde solutions are presented in Table \ref{tab_orbitVanth}, where the prograde presents 
smaller $\chi^2$ per degree of freedom ($\chi^2_{\rm pdf}$)\footnote{The much smaller uncertainties on the orbital parameters of the retrograde solution, when compared to the prograde solution, is an indication that the solution was trapped in a local minimum, i.e. any deviation from that value results in a degenerate fit, as can be noticed in Figure \ref{fig:chiVanthRetro}, top left panel.}. 
The 2014 point has a residual of almost 30 mas, which is partially responsible for a $\chi^2_{\rm pdf}$ significantly larger than the unity and may be caused by either a systematic error in Orcus' ephemeris, a timing issue on the data, a non-circular apparent limb, or, most probably, a combination of these. The low uncertainties of the 2006 HST data points and 2016 observations from ALMA are also contributing to the increase in the $\chi^2_{\rm pdf}$.  
All this data obtained from different sources, imaging, stellar occultation and millimetres, have proved to be difficult to analyse in a consistent way, as they may have different sources of systematic errors.
The $\chi^2$ plots for some orbital elements are provided in Appendix \ref{chi2plots}.

\begin{figure}
    \centering
    \includegraphics[width=0.7\linewidth]{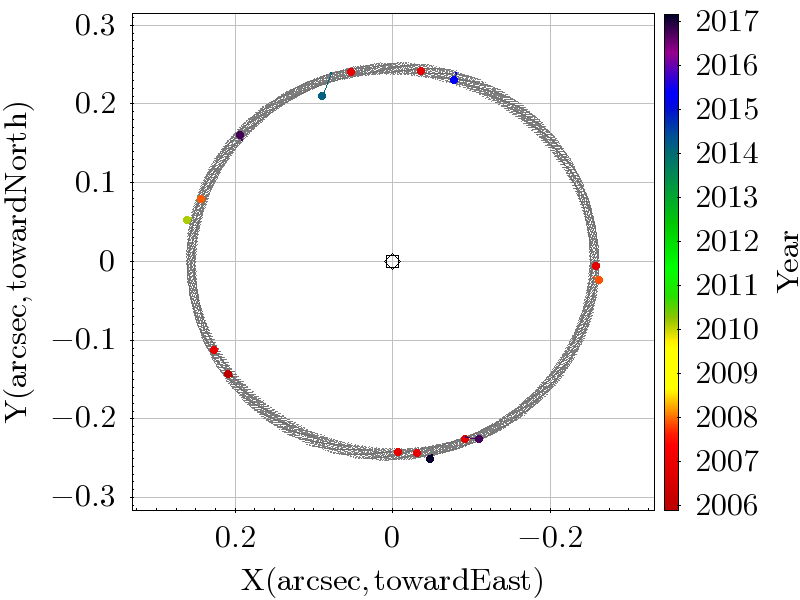}
    \caption{ Vanth's orbit around Orcus projected in the sky plane. The data points are coloured according to the observation year. 
    The calculated orbit is indicated by grey points, and its thickness stems from the changing viewing geometry between 2005 and 2017.
    The small segments connect the observed positions to the calculated solution. Even if the 2014 occultation point (in purple) has been corrected by the photocentric shift, it is still off the orbit solution, augmenting the $\chi^2_{\rm pdf}$ value of the fit.}
    \label{fig:VanthOrbitpro}
\end{figure}

\begin{figure}
    \centering
    \includegraphics[width=0.7\linewidth]{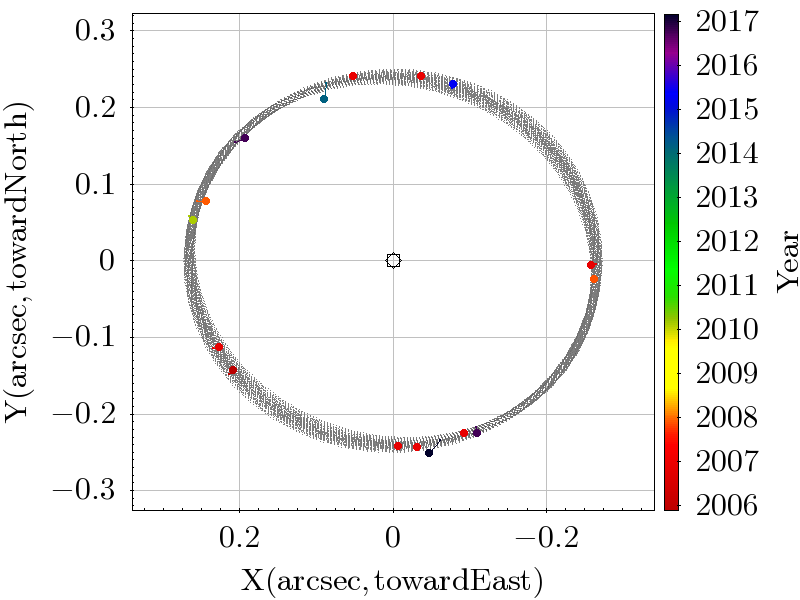}
    \caption{ Similar as Figure \ref{fig:VanthOrbitpro} but for the retrograde solution of Vanth's orbit around Orcus projected in the sky plane.}
    \label{fig:VanthOrbitRetro}
\end{figure}

These results can be compared with previously published values \cite{Brown18,Carry11,Grundy19,Proudfoot24}.
The retrograde solution is comparable to the solutions presented by \cite{Grundy19}. Both solutions are consistent with a circular motion. If we assume the orbit became circularized, this could have been caused by a physical mechanism, such as tidal forces, friction with particles composing a dust ring, possibly during the satellite's own accretion, or gravitational interaction with another (unknown) satellite.
Differently from \cite{Proudfoot24}, our non-Keplerian solution does not improve the quality of the fit, so we do not fit the $J_2$ factor. 
This difference may come from the algorithms' different capabilities for exploring the space parameters.  Figure~\ref{fig:VanthOrbitpro} presents the prograde orbit of Vanth around Orcus projected in the sky plane, while Figure~\ref{fig:VanthOrbitRetro} presents the retrograde orbit of Vanth.

\begin{table}[ht]
%    \centering
    \begin{tabular}{l l l| l l l | l l}
       \multicolumn{8}{c}{Vanth (Orcus/1)} \\
       \hline \hline
\multicolumn{3}{l|}{ Orbital elements EQJ2000 } & 
\multicolumn{3}{l|}{ Derived parameters } &
\multicolumn{2}{l}{Observing data set} \\ 
\hline \hline
       \multicolumn{8}{c}{\textbf{Prograde}} \\
       \hline

P (days)  & 9.53894 & $\pm$0.00010  &  \multicolumn{1}{c}{Syst. mass} & \multirow{2}{*}{6.163} & \multirow{2}{*}{$\pm$0.079} & N. Obs. & 14 \\

a (km) & 8912 & $\pm$38 & (x10$^{20}$~kg) &  &  & N. Occ. & 2   \\

e   & 0.0010  & < 0.0030 & $\alpha_{\rm p}$  ($^\circ$) & 327.0 & $\pm$2.1 & Time span &  4132 days   \\

i ($^\circ$)  & 66.4 & $\pm$1.8 &  $\delta_{\rm p}$ ($^\circ$)  & 23.62 & $\pm$0.85 & 1$^{\rm st}$ date & 2005-11-12 \\

$\Omega$ ($^\circ$)   & 57.0 & $\pm$2.1 & $\lambda_{\rm p}$ ($^\circ$) & 338.3 & $\pm$2.2 & RMS (mas) & 6.9 \\

$\omega$ ($^\circ$)   & 158.4 & $\pm$7.0 & $\beta_{\rm p}$ ($^\circ$) & 34.20  & $\pm$0.92 & $\chi^2_{\rm pdf}$& 7.0  \\

$t_{\rm p}$ (JD) & 2455994.27 & $\pm$0.19 & & & & &  \\
       \hline   \hline
              \multicolumn{8}{c}{\textbf{Retrograde}} \\
       \hline

P (days)  & 9.5391320 & $\pm$0.00008  &  \multicolumn{1}{c}{Syst. mass} & \multirow{2}{*}{6.92413} & \multirow{2}{*}{$\pm$0.00035} & N. Obs. & 14 \\

a (km) &  9264.33 & $\pm$0.13 & (x10$^{20}$~kg) &  &  & N. Occ. & 2   \\

e   & 0.0000  & < 0.00073 & $\alpha_{\rm p}$  ($^\circ$) & 323.9 & $\pm$2.0 & Time span &  4132 days   \\

i ($^\circ$)  & 109.13 & $\pm$0.63 &  $\delta_{\rm p}$ ($^\circ$)  & -19.13 & $\pm$0.63 & 1$^{\rm st}$ date & 2005-11-12 \\

$\Omega$ ($^\circ$)   & 53.9 & $\pm$2.0 & $\lambda_{\rm p}$ ($^\circ$) & 320.0 & $\pm$1.8 & RMS (mas) & 7.3 \\

$\omega$ ($^\circ$)   & 126.3 & $\pm$1.9 & $\beta_{\rm p}$ ($^\circ$) & -5.00	  & $\pm$0.69 & $\chi^2_{\rm pdf}$& 15.5  \\

$t_{\rm p}$ (JD) & 2455993.450 & $\pm$0.048 & & & & &  \\
\hline
    \end{tabular}
    \caption{Vanth's orbital elements centred at Orcus, expressed in Equatorial J2000: orbital period $P$, semi-major axis $a$, eccentricity $e$, inclination $i$, the longitude of the ascending node $\Omega$, the argument of pericentre $\omega$, time of passage at pericentre  $t_{\rm p}$ in Julian date. 
    The system total mass M, the  Equatorial J2000  coordinates ($\alpha_{\rm p},\delta_{\rm p}$) and the  Ecliptic J2000  coordinates of the orbital pole ($\lambda_{\rm p},\beta_{\rm p}$) are provided. 
    Finally, the number of relative positions from direct images and the number of positions from stellar occultations, the period between the observations, the date of the first used observation, and the dispersion of the data with respect to the fitted orbit are also provided. Uncertainties are given at 1$\sigma$.}
    \label{tab_orbitVanth}
\end{table}

\subsection{Weywot (Quaoar/1)}
\label{subsec:Weywot}

Quaoar is one of the largest Trans-Neptunian Objects orbiting in the Hot Classicals region with a semi-major axis of 43.5 au, a small eccentricity of 0.041, and an orbital inclination of 8.0 degrees. It has an equivalent diameter of 1110~$\pm$~10~km, measured with a stellar occultation in 2011 under the assumption of a Maclaurian shape \cite{Braga-Ribas2013}. In 2023, two rings were announced around Quaoar \cite{Morgado2023,Pereira2023} at distances of about  2500~km and 4000~km, well outside the classical Roche limit,  estimated to be at less than 1800~km from Quaoar's centre \cite{Morgado2023}. 
The outermost ring is inhomogeneous and very close to the 1/3 spin-orbit resonance with Quaoar, considering the rotation period of $\sim$17.7~hours, which is the case for a triaxial object \cite{Ortiz03}. 

Its moon, Weywot, was found in 2006 \cite{BrownSuer07}, and had its diameter estimated to be $81~\pm~11$~km,  based on thermal data, under the assumption of equal albedo relative to Quaoar of 12~$\pm$~1~\% \cite{Fornasier13}.  Its orbital semi-axis is on the order of 13300~km with a period of $\sim$12.4~days.

Several attempts to detect stellar occultations by Weywot have been made by our team since 2012, but the first detection made in 2019 was a lucky catch, with both Quaoar and Weywot in the same light curve. With the preliminary relative position obtained from that event, we used GENOID to update the orbital fit and calculated its ephemeris, which was available through IMCCE's Miriade service. 

Using NIMA ephemeris for Quaoar and GENOID ephemeris for the relative position of Weywot, we predicted and successfully detected occultations by Weywot in June 2022 and 2023, crossing the USA, as well as an event in May 2023, crossing New Zealand. Although the detailed analysis of these events is the subject of a work in development, all of them are consistent with a circular limb fit with a diameter of $\sim$170~km, significantly larger than the thermally-derived value $81 \pm 11$~km mentioned earlier \cite{Fornasier13},   which indicates a much darker surface when compared to Quaoar's.    In this work, we assume a spherical shape with a 170-km diameter to derive Weywot's position relative to Quaoar. 
 
Quaoar's NIMAv19 ephemeris is based on astrometric positions obtained from about 20 stellar occultations detected from 2011 up to 2024, as well as MPC and proprietary classical astrometric positions. Due to the much better accuracy and weight given to the occultation derived positions,
the ephemeris is dominated by these positions, thus, it is not affected by the photocentre displacement cause by the satellite present on the direct unresolved image. Therefore we do not correct Weywot's relative positions reported here from the photocentre offset, as NIMAv19 already provides the primary body centre of figure position.

\subsubsection{August 04, 2019}

An occultation by Quaoar was predicted to cross Namibia on August 04, 2019. Observations were attempted from the Tivoli site. Thankfully, the observer started to acquire data fourteen minutes before the predicted time, registering the occultation by Weywot eleven minutes before the Quaoar event. 

Quaoar's chord had a length of  1121~$\pm$~15~km; thus, it is very close to being diametric. From this observation, we derived the astrometric position of Quaoar presented in table \ref{tab_Weywotpos2019}. 
 Using the diameter of 170~km (as explained above), we can fit two circles to the single 168~$\pm$~15~km observed chord. One solution has the circle's centre to the North of the chord, and the other has the centre to the South of it. Both are presented in Table \ref{tab_Weywotpos2019}. Considering the occulted star, main body, and chord fit, the final relative position $(X,Y)$ and uncertainties are given in Appendix \ref{app_positions}, table \ref{tab_posWeywot}.

\begin{table}[ht]
    \centering
    \begin{tabular}{c c c c c} 

    \multicolumn{5}{c}{Weywot: August 04, 2019 astrometry}  \\

    \hline \hline
    
    \multirow{2}{*}{Solution}   & 
    Date and Time (UT)  &
    Right ascension (h m s)    &  
     Offset &
    Uncertainty \\ 
    
     &    
     Julian date &
     Declination  ($^\circ$ ’ ”)  &  
     $ (X,Y)$ (km) &
     (km)      \\
     \hline \hline
 \multicolumn{5}{c}{\bf{Quaoar}}  \\
\hline
\multirow{2}{*}{South}   &
    2019-08-04 17:24:52.560 &
    18 05 34.78680  $\pm$ 0.37 &
    -22 &
    10 \\
    
      &
     2458700.2256083335 &
    -15 21 02.068139 $\pm$ 1.38 &
    -19 &
    42 \\
     \hline 

    \hline \hline
    \multicolumn{5}{c}{\bf{Weywot}}  \\
\hline

    \multirow{2}{*}{North}   &
    2019-08-04 17:24:52.560 &
    18 05 34.75794 $\pm$ 0.27 &
     -12761 &
    8 \\

      &
     2458700.2256083335 &
    -15 21 02.1335 $\pm$ 0.60 &
     -1977 &
    18 \\
     \hline 

    \multirow{2}{*}{South}   &
    2019-08-04 17:24:52.560 &
    18 05 34.75795 $\pm$ 0.25 &
     -12757 &
    7 \\

      &
     2458700.2256083335 &
    -15 21 02.1347 $\pm$ 0.62 &
     -2013 &
    19 \\
    \hline
    \end{tabular}
    \caption{Similar to Table~\ref{tab_Vanthpos2014}, Weywot position relative to Quaoar for each possible solution, i.e., centre to the North or South of the chord. The Quaoar centre was obtained from the limb fitting of the positive chord,
    considering the propagated Gaia DR3 star position RA: 18h 05m 34.78660s $\pm$ 0.13 mas, DEC: -15$^\circ$ 21’ 02”.04630 $\pm$ 0.14 mas. 
    Weywot's ICRS and $ (X,Y)$ position are relative to the obtained Quaoar's centre, not considering the main body uncertainty. Different from  Table \ref{tab_posWeywot}, here,  the 1$\sigma$ uncertainties on $(X,Y)$ consider only  the expected satellite diameter fitted to the chord.
    The object was at a distance of 42.08 au.
    }
    \label{tab_Weywotpos2019}
\end{table}

\subsubsection{June 11, 2022}

Using GENOID and NIMA updated ephemeris, which considered the former occultation, this event was predicted to cross the USA. The 
 Occultation Timing Association (IOTA)  amateur community was alerted, and observations were also made from Flagstaff facilities. Two positive and one close negative chords were detected on Weywot. Quaoar's shadow crossed the Amazon forest, but no observations could be made.  

Although a 20\% smaller circle could be used, we fitted a 170~km circle to calculate Weywot's astrometric position (see table \ref{tab_Weywotpos22}). NIMAv19 ephemeris uses astrometric positions from several stellar occultations; thus, the uncertainty on its position is of only a couple of mas, i.e., less than 60~km at Quaoar's distance. So, we used NIMAv19 Quaoar's position, not applying any photocentre correction, to calculate Weywot's relative position given in Appendix \ref{app_positions}, Table~\ref{tab_posWeywot},  where the uncertainty on Quaoar position is considered, as well as the uncertainties stemming from the star position and the circular fit to the chords.

\begin{table}[ht]
    \centering
    \begin{tabular}{c c c r c} 

    \multicolumn{5}{c}{Weywot: June 11, 2022 astrometry}  \\

    \hline \hline
    
    \multirow{2}{*}{Solution}   & 
    Date and Time (UT)  &
    Right ascension (h m s)    &  
    Offset &
    Uncertainty \\ 
    
     &    
     Julian date &
     Declination  ($^\circ$ ’ ”)  &  
     $(X,Y)$ (km) &
     (km)      \\
     \hline \hline

    \multirow{2}{*}{Circle}   &
    2022-06-11 09:13:52.920  &
    18 26 06.65650 $\pm$ 0.15 &
     -6598.2 &
    1.6 \\

      &
     2459741.8846402778 &
    -15 07 47.33584 $\pm$ 0.15 &
     -3870.2 &
    1.5 \\
     \hline 

    \end{tabular}
    \caption{Similar to  Table~\ref{tab_Vanthpos2014},using the NIMAv19 ephemeris and assuming that Quaoar is centred,
    considering the propagated Gaia DR3 star position RA: 18h 26m 06.67152s $\pm$ 0.14 mas, DEC: -15$^\circ$ 07’ 47”.49461 $\pm$ 0.14 mas. 
    The 1$\sigma$ uncertainties on $(X,Y)$ consider Weywot's limb fit only. The object was at a distance of 41.785 au.
    }
    \label{tab_Weywotpos22}
\end{table}

\subsubsection{May 26, 2023}

Using the same updated ephemeris as for the 2022 event, we predicted the occultation of Weywot, followed by Quaoar on May 26, 2023, crossing New Zealand and South Australia. Weywot and Quaoar events were detected from the northmost site in Blenheim/NZ, while Quaoar's was also detected from the other three stations in New Zealand.

The four chords detected on Quaoar allow obtaining a precise astrometric position (Table \ref{tab_WeywotposMay23}) and thus calculate Weywot's relative position. With a chord length of 180~$\pm$~19~km, due to the uncertainty, two solutions are possible for the object's centre, which are presented in Table \ref{tab_WeywotposMay23}. Considering the circular fit and the uncertainty on Quaoar's position, the relative offset position is given in Appendix \ref{app_positions}, Table \ref{tab_posWeywot}.

\begin{table}[ht]
    \centering
    \begin{tabular}{c c c r c} 

    \multicolumn{5}{c}{Weywot: May 26, 2023 astrometry}  \\

    \hline \hline
    
    \multirow{2}{*}{Solution}   & 
    Date and Time (UT)  &
    Right ascension (h m s)    &  
    Offset &
    Uncertainty \\ 
    
     &    
     Julian date &
     Declination  ($^\circ$ ’ ”)  &  
     $(X,Y)$ (km) &
     (km)      \\
     \hline \hline
    \multicolumn{5}{c}{\bf Quaoar}  \\
\hline
    \multirow{2}{*}{Elliptical}   &
    2023-05-26 15:53:06.640  &
    18 32 44.928582 $\pm$ 0.20 &
    49.2  &
    1.4 \\

      &
     2460091.1618824075 &
    -15 03 49.081808 $\pm$ 0.33 &
    11.1 &
    8.0 \\
     \hline \hline
     
    \multicolumn{5}{c}{\bf Weywot}  \\
\hline
    \multirow{2}{*}{North}   &
    2023-05-26 15:53:06.640  &
    18 32 44.90169 $\pm$ 0.30 &
     -11829.4 &
    7.0 \\

      &
     2460091.1618824075 &
    -15 03 49.0467 $\pm$ 0.48 &
     1066.4 &
    15.5 \\ 
    \hline 
    
    \multirow{2}{*}{South}   &
    2023-05-26 15:53:06.640  &
    18 32 44.90169 $\pm$ 0.30 &
     -11829.8 &
    7.0 \\

      &
     2460091.1618824075 &
    -15 03 49.0476 $\pm$ 0.48 &
     1039.7 &
    15.7 \\
 
     \hline 

    \end{tabular}
    \caption{Similar to  Table~\ref{tab_Vanthpos2014}, Weywot position relative to Quaoar. 
    A Quaoar centre was obtained from the limb fitting of the four positive chords. The ICRS positions
    consider the propagated Gaia DR3 star position RA: 18h 32m 44.92889s $\pm$~0.19 mas, DEC: -15$^\circ$ 03’ 48”.97855 $\pm$~0.19 mas. 
    Weywot's position $(X,Y)$ relative to Quaoar's centre, does not consider the main body uncertainty. The object was at a  geocentric  distance of 41.8730 au.
    }
    \label{tab_WeywotposMay23}
\end{table}

\subsubsection{June 22, 2023}

The occultation of June 22, 2023, was very similar to that of 2022 and was also detected by IOTA astronomers and Flagstaff facilities. Five positive chords were recorded, and close negatives to the North and South limited the possible solutions.  A 170~km circle is compatible with the observed chords, so we fitted the circular limb to derive Weywot's astrometric position (Table \ref{tab_WeywotposJune23}). NIMAv19 ephemeris was used to calculate the relative position available in Table \ref{tab_posWeywot}, where the uncertainty on Quaoar position is considered, as well as the star position and the circular fit to the chords.

\begin{table}[ht]
    \centering
    \begin{tabular}{c c c r c} 

    \multicolumn{5}{c}{Weywot: June 22, 2023 astrometry}  \\

    \hline \hline
    
    \multirow{2}{*}{Solution}   & 
    Date and Time (UT)  &
    Right ascension (h m s)    &  
    Offset &
    Uncertainty \\ 
    
     &    
     Julian date &
     Declination  ($^\circ$ ’ ”)  &  
     $(X,Y)$ (km) &
     (km)      \\
     \hline \hline

    \multirow{2}{*}{Diametrical}   &
    2023-06-22 08:00:48.900  &
    18 30 45.75455 $\pm$ 0.35 &
     -11956.1 &
    0.1 \\

      &
     2460117.8338993057 &
    -15 03 33.491764 $\pm$ 0.39 &
     -3420.1 &
    0.8 \\
     \hline 

    \end{tabular}
    \caption{Similar to  Table~\ref{tab_Vanthpos2014},  Weywot position relative to Quaoar
    using the NIMAv19 ephemeris and assuming that Quaoar is centred.
    Considering the propagated Gaia DR3 star position RA: 18h 30m 45.78240s $\pm$ 0.35 mas, DEC: -15$^\circ$ 03’ 33”.64224 $\pm$ 0.39 mas. The object was at a geocentric distance of 41.7151 au.
    The uncertainties on $(X,Y)$ only consider the fit to the object limb.
    }
    \label{tab_WeywotposJune23}
\end{table}

\subsubsection{Orbital elements}

Using the four relative positions derived from the detected stellar occultations and nine direct images (see table \ref{tab_posWeywot}), we calculated Weywot's orbit relative to Quaoar. Using the Keplerian model, we obtained the solution presented in Table~\ref{tab_orbitWeywot}, where the system mass and the orbital pole are also shown.  These results represent a significant improvement in terms of uncertainties from previous works \cite{Vachier12,Proudfoot24}, with smaller mass and eccentricity and larger semi-major axis. The 2011 observations made with Keck \cite{Fraser13} present systematic negative differences with respect to our solution, with values that reach 14 mas. This may explain the significant differences between our solution and those that do not rely on occultation-derived relative positions.

\begin{figure}
    \centering
    \includegraphics[width=0.7\linewidth]{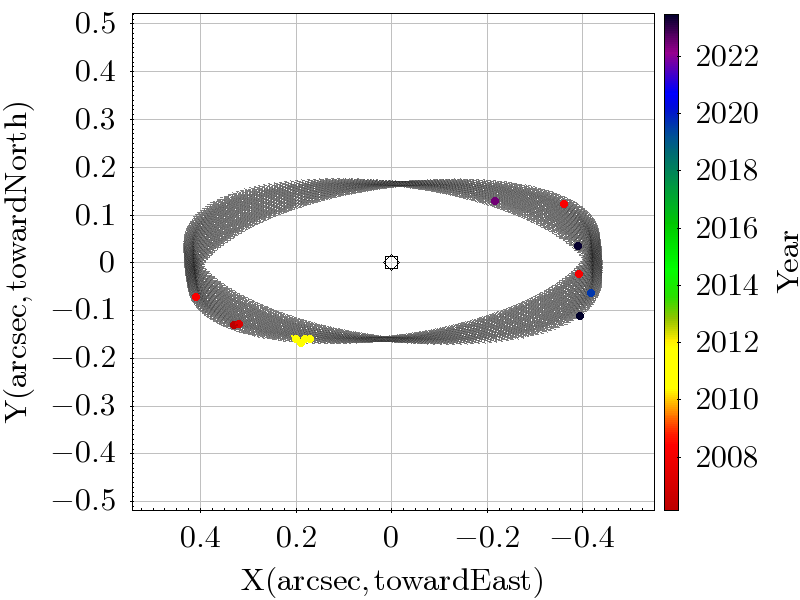}
    \caption{ The same as Fig.~\ref{fig:VanthOrbitpro} for Weywot's orbit around Quaoar. The grey points are now significantly spread out due to the changing viewing geometry of Weywot's orbit between 2006 and 2023.}
    \label{fig:WeywotOrbit}
\end{figure}

The Figure~\ref{fig:WeywotOrbit} displays Weywot's orbit around Quaoar.
The $\chi^2$  dependence on some orbital elements can be seen in Appendix \ref{chi2plots}, Figure \ref{fig:chiWeywot}. 

The mass, period, and eccentricity are fundamental parameters that eventually inform us about the density of the central body and the possible dynamical link between Weywot and Quaoar's rings. We note in particular that the currently measured pole of Quaoar's main ring Q1R ($\alpha_{\rm Q1R}= 259.8^\circ \pm 0.2^\circ$, $\delta_{\rm Q1R}= 53.5^\circ \pm 0.3^\circ$ \cite{Pereira2023}) presents a difference of $4.8^\circ \pm 1.6^\circ$ relative to the Weywot's orbital pole given in Table~\ref{tab_orbitWeywot}, suggesting a possible mutual inclination between Weywot and Q1R. Nevertheless, it is important to consider that this may come from an unknown systematic error on the ring pole determination or from the assumption that it is circular and equatorial to Quaoar.

\begin{table}[ht]
%    \centering
% 044.cerfeuil.kepler.Grundy_newobs2023_noHST
    \begin{tabular}{l l l| l l l | l l}
       \multicolumn{8}{c}{Weywot (Quaoar/1)} \\
       \hline \hline
\multicolumn{3}{l|}{ Orbital elements EQJ2000 } & 
\multicolumn{3}{l|}{ Derived parameters } &
\multicolumn{2}{l}{Observing data set} \\ 
\hline 

P (days)  & 12.431013 & $\pm$0.00021  &  \multicolumn{1}{c}{Syst. mass} & \multirow{2}{*}{1.208} & \multirow{2}{*}{$\pm$0.063} & N. Obs. & 9 \\

a (km) & 13309 & $\pm$231 & (x10$^{21}$~kg) &  &  & N. Occ. & 4   \\

e   & 0.018 & $\pm$0.016 & $\alpha_{\rm p}$  ($^\circ$) & 266.9 & $\pm$1.5 & Time span & 6336 days   \\

i ($^\circ$)  & 38.04 & $\pm$1.4 &  $\delta_{\rm p}$ ($^\circ$)  & 51.9 & $\pm$1.4 & 1$^{\rm st}$ date & 2006-02-14 \\

$\Omega$ ($^\circ$)   & 356.9 & $\pm$1.5 & $\lambda_{\rm p}$ ($^\circ$) & 262.8 & $\pm$3.3 & RMS (mas) & 4.3 \\

$\omega$ ($^\circ$)   & 292 & $\pm$25 & $\beta_{\rm p}$ ($^\circ$) & 74.6 & $\pm$1.4 &  $\chi^2_{\rm pdf}$&  0.47  \\

$t_{\rm p}$ (JD) & 2453780.42 & $\pm$0.82 & & & & & \\
       \hline        
    \end{tabular}
    \caption{Same as Table \ref{tab_orbitVanth} but for Quaoar's satellite, Weywot. Uncertainties are given at 1$\sigma$}
    \label{tab_orbitWeywot}
\end{table}

\section{Discussion and Future Prospects}
\label{sec_discuss}

The study of the Outer Solar System using stellar occultation started with observing Pluto events in the 80's (see \cite{Elliot1989,Brosch1995} and the review by \cite{Sicardy2024}), but it has been generally applied to other small bodies since 2009 \cite{Elliot10}. Since then, numerous interesting results have allowed a deeper understanding of this region \cite{Sicardy2024}. 

In the era of LSST (Legacy Survey of Space and Time)  and the Extremely Large Telescopes, the field will make another significant step toward studying the Transneptunian Binaries.
The LSST will discover and repeatedly observe tens of thousands of TNOs, up to $\sim$24th magnitude, providing accurate astrometric positions. This will lead to better orbit computations, allowing reliable predictions of stellar occultations and enabling an efficient occultation survey for the search of TNOBs on smaller objects.
The future extremely large telescopes will be capable of providing direct images with resolutions better than ten milli-arcseconds (mas) with a contrast of 10$^{-8}$ in flux between quite close different sources. This will allow more observations of the TNO satellites, providing further constraints on their composition and rotation period. It will also offer relative positions that will be used to update their orbits, allowing better predictions of occultations by the satellites\footnote{The LIneA Occultation Prediction Database was built to provide predictions of all Small Solar System Objects in the LSST era: \url{https://solarsystem.linea.org.br/}.}.

Stellar occultations now have the capacity to provide km-level sensitivities for both the shape and the on-sky positions of small objects of the outer Solar System \cite{Sicardy2024}. They also allow us to probe the surroundings of these objects, especially the Cold Classicals, searching for small satellites or binary pairs. 
Using the Earth as a big telescope, thanks to the Pro-Am collaborative work, it will be possible to  perform surveys for determining  better the number of binary objects below the resolving power of the current telescopes ($\sim$~1200~km), which will be an essential step in understanding solar system formation.

 The probability of detecting a pair of objects will depend on their sizes, relative distance, prediction uncertainty and number of telescopes observing the event. Considering that LSST single epoch images will typically deliver astrometric positions with uncertainties of 10 mas, the TNO ephemeris will have similar precision on short-term predictions (i.e., a few years) \cite{Camargo18}. 
 %This is better than resolving the power limit. 
 Let us consider that we will want to use an occultation to scan about two thousand kilometres to probe the high-resolution images' blind region, plus the prediction uncertainty. If each component of the binary system have 50~km in diameter, we will want a fence like distribution perpendicular to the shadow path of, at least, one telescope every 25~km, to guarantee the detection. This corresponds to a total of 80 telescopes along the 2000~km region. This may seem to be a large number, but when observing campaigns can count on the collaboration of the amateur community, a large number of telescopes can be involved, as has been the case on the occultations by Triton \cite{MarquesOliveira22} and 2002~MS$_4$ \cite{Rommel2023}. 

Regarding the satellites, the new Extremely Large telescopes with direct imaging power of a few mas \cite{TMT2022}, that is, a few hundred kilometres at the TNO distance, will enable more observations of the close environment of the big TNOs and thus their satellites. With more relative positions, better orbits and ephemerides will be calculated. 

We have shown that with GENOID plus NIMA, we can provide reliable predictions of stellar occultations by TNO moons  as soon as relative positions with sufficiently good quality are available. With the detection of these events from multiple sites, their physical properties can be obtained. As explained in Section \ref{sec:Introduction}, information such as shape, albedo, and density allows for a better understanding of their formation history, constraining the Solar System's dynamical evolution scenarios. 

GENOID allows for a complete exploration of the orbital space parameters, finding Keplerian solutions for most systems. It also allows studying the influence of the main body's gravitational field perturbations caused by its oblateness and the mutual influence of multiple satellites in the system. A better ephemeris enables the improvement of dynamical studies on resonances, tides, and implications for ring dynamics and confinement \cite{Sicardy19}. 

We found that when only the satellite is detected during a stellar occultation, and the main body's ephemeris is based mainly on direct images, it is important to correct its relative position from the photocentric offset caused by the satellite. Conversely, in the case of Quaoar, the photocentre offset has no influence, as its ephemeris is dominated by positions derived from previous stellar occultations.

 We have shown that the Orcus-Vanth and Quaoar-Weywot orbital solutions are consistent with purely Keplerian motions. New orbital parameters and masses were obtained, fundamental information for the study of the dynamical environment (e.g. search for stable regions) and for determining the central object's physical properties, such as its density. The new ephemeris will allow for better prediction of future stellar occultations for the full characterization of these satellites and possibly better constraining their origin scenarios.

 This work presents a list of all the TNO satellites (excepted Charon) that were successfully detected so far using stellar occultations, and discusses the implications of this starting field of research. It will undoubtedly grow and provide significant results, allowing us to understand better the history of solar system formation.

%%%%%%%%%%%%%%%%%%%%%%%%%%%%%%%%%%%%%%%%%%%%%%%%%%%%%%%%%%%%%%%%%%%%%%%%%%%%%%%%%
%%%%%%%%%%%%%%% End of main text %%%%%%%%%%%%%%%%%%%%%%%%%%%%%%
%%%%%%%%%%%%%%

%\section{Conclusion}
%\label{sec:Conclusion}

\ack{This study was financed in part by the Coordenação de Aperfeiçoamento de Pessoal de Nível Superior - Brasil (CAPES) - Finance Code 001 and National Institute of Science and Technology of the e-Universe project (INCT do e-Universo, CNPq grant 465376/2014-2). FBR acknowledges CNPq grant 316604/2023-2. GM acknowledges the CAPES scholarship number 88887.705245/2022-00. The ephemeris calculations of the position of the bodies of the Solar System were carried out by the IMCCE ephemeris calculation service through its Solar System portal (\url{https://ssp.imcce.fr}). We thank the observers who reported the observations used in this work and are co-authors of the respective works, yet to be published, with the complete analysis of these events.}

\begin{appendices}

\section{Satellite positions}
\label{app_positions}

The positions taken from the literature used on our orbital fit with GENOID are given in Table \ref{tab_posVanth} for Vanth (Orcus/1) and Table \ref{tab_posWeywot} for Weywot (Quaoar/1).

\begin{table}[ht]
    \centering
    \begin{tabular}{c l l l l c }
\multicolumn{6}{c}{ Vanth  offset  positions relative to Orcus} \\
\hline \hline
 Observation Date & \multicolumn{1}{c}{ $X$ ('')} & \multicolumn{1}{c}{ $Y$ ('')} &  $X_{\rm err}$ ('') &  $Y_{\rm err}$ ('') & Source (Ref.) \\
\hline
 2005-11-13T03:56:09.600 &  0.20792  & -0.14342  & 0.00492 & 0.00992 & HST \cite{Brown10}    \\
 2006-10-31T20:51:21.600 &  0.22597  & -0.11307  & 0.00102 & 0.00071 & HST \cite{Brown10}     \\
 2006-11-03T01:36:14.184 & -0.09310  & -0.22635  & 0.00050 & 0.00050 & HST \cite{Brown10}     \\
 2006-11-04T20:47:02.400 & -0.25893  & -0.00560  & 0.00050 & 0.00079 & HST \cite{Brown10}     \\
 2006-11-12T01:53:45.600 & -0.00694  & -0.24344  & 0.00185 & 0.00050 & HST \cite{Brown10}     \\
 2006-11-16T14:08:09.600 & -0.03718  &  0.24113  & 0.00058 & 0.00050 & HST   \cite{Brown10}     \\
 2006-11-26T15:30:14.400 &  0.05228  &  0.24033  & 0.00085 & 0.00050 & HST   \cite{Brown10}     \\
 2006-12-10T20:03:50.400 & -0.03180  & -0.24410  & 0.00050 & 0.00050 & HST   \cite{Brown10}     \\
 2007-11-11T19:01:55.200 & -0.26300  & -0.02400  & 0.00500 & 0.00500 & HST   \cite{Brown10}     \\
 2007-12-05T06:43:12.000 &  0.24288  &  0.07814  & 0.00241 & 0.00418 & HST   \cite{Brown10}     \\
 2010-02-23T05:41:04.704 &  0.26000  &  0.05200  & 0.02500 & 0.02500 & VLT   \cite{Carry11}      \\
 2014-03-01T16:19:01.120 &  0.08891  &  0.21006  &  0.0082  &  0.0088  & OCC N.$^\dag$      \\
 2014-03-01T16:19:01.120 &  0.08504  &  0.20690  &  0.0083  &  0.0136  & OCC S.$^\dag$       \\
 2015-04-05T09:07:12.108 & -0.07854  &  0.22974  & 0.00100 & 0.00100 & Keck  \cite{Grundy19}         \\
 2016-10-11T11:22:40.890 &  0.1926   &  0.1596   & 0.005   & 0.005   & ALMA  \cite{Brown23}	       \\
 2016-10-15T12:05:15.500 & -0.1095   & -0.2264   & 0.005   & 0.005   & ALMA  \cite{Brown23}	      \\
 2017-03-07T06:53:51.000 & -0.0475   & -0.2521   & 0.011   & 0.012   & OCC   \cite{Sickafoose19}$^\dag$ \\
\hline
 \end{tabular}
    \caption{
     Vanth's on-sky offset $(X,Y)$ relative to Orcus and their estimated errors, in arcsec.
    % to Vanth relative positions in the  celestial  cartesian plane, with values and uncertainties given in seconds of arc. 
    Occultation-derived (OCC) positions obtained in this work, corrected for the photocentre offset, are marked with '$^\dag$'. The final uncertainty in the quadratic sum of the uncertainties of the Orcus position with the stellar position and the Vanth's radius ($222 \pm 5$~km \cite{Sickafoose19}) to account for its unknown shape.
    }
    \label{tab_posVanth}
\end{table}

%%%%%%%%%%%%%%%%%%%%%%%%%%%%%%%%%%%%%%%%%%%%%%%%%%%

\begin{table}[h]
    \centering
    \begin{tabular}{c l l l l c }
\multicolumn{6}{c}{Weywot  offset  positions relative to Quaoar} \\
\hline \hline
 Observation Date & \multicolumn{1}{c}{  $X$ ('')} & \multicolumn{1}{c}{ $Y$('')} &  $X_{\rm err}$('') &  $Y_{\rm err}$('') & Source (Ref.) \\
\hline
 2006-02-14T21:30:32.940 &  0.3288 & -0.1311 & 0.00797 & 0.00253 &  HST  \cite{Fraser13} \\
 2007-03-19T15:22:58.512 &  0.3177 & -0.1288 & 0.00188 & 0.00264 &  HST  \cite{Fraser13} \\
 2008-03-10T04:55:31.980 & -0.3922 & -0.0251 & 0.00396 & 0.00363 &  HST  \cite{Fraser13} \\
 2008-03-15T01:31:01.992 &  0.4082 & -0.0721 & 0.00371 & 0.00144 &  HST  \cite{Fraser13} \\
 2008-03-20T16:28:01.992 & -0.3622 &  0.1230 & 0.00332 & 0.00370 &  HST  \cite{Fraser13} \\
 2011-06-07T07:53:07.584 &  0.1700 & -0.1600 & 0.01000 & 0.01000 &  Keck \cite{Fraser13} \\
 2011-06-07T09:32:12.768 &  0.1800 & -0.1600 & 0.01000 & 0.01000 &  Keck \cite{Fraser13} \\
 2011-06-07T10:58:56.640 &  0.1900 & -0.1700 & 0.01000 & 0.01000 &  Keck \cite{Fraser13} \\
 2011-06-07T12:25:19.776 &  0.2000 & -0.1600 & 0.01000 & 0.01000 &  Keck \cite{Fraser13} \\
 2019-08-04T17:24:52.560 & -0.41820 & -0.06476 &  0.0028  &  0.0032  &  OCC N.$^\dag$      \\
 2019-08-04T17:24:52.560 & -0.41797 & -0.06596 &  0.0028  &  0.0031  &  OCC S.$^\dag$      \\
 2022-06-11T09:13:52.920 & -0.2177 &  0.1277 &  0.0035  &  0.0031  &  OCC $^\dag$      \\ 
 2023-05-26T15:53:06.640 & -0.38953 &  0.03512 &  0.0028  &  0.0029  &  OCC N.$^\dag$      \\
 2023-05-26T15:53:06.640 & -0.38952 &  0.03424 &  0.0028  &  0.0029  &  OCC S.$^\dag$      \\
 2023-06-22T08:00:48.900 & -0.3952 & -0.1130 &  0.0035  &  0.0030  &  OCC $^\dag$      \\
\hline
 \end{tabular}
    \caption{
     The same as Table~\ref{tab_posVanth} for Weywot's position relative to Quaoar.
    Detailed analyses of the stellar occultation results will be presented in future work. The  June 2022 and 2023 results are based on the NIMAv19 position of Quaoar, which has an estimated uncertainty of $\sigma_\alpha(\cos\delta)=2.0$~mas~and~$\sigma_\delta=1.0$~mas~for both events.  The final uncertainty is the quadratic sum of Quaoar's position with the stellar position and the Weywot's radius (85~km)(assumed spherical), to account for its unknown shape.
    Conversely, the 2019 and May 2023 events have smaller uncertainties, as Quaoar was also detected, providing more accurate relative positions.}
    \label{tab_posWeywot}
\end{table}

\section{Fitness function}
\label{chi2plots}

GENOID is a statistical method for minimizing the discrepancies between observed data and the theoretical model. Like the least squares method, this approach is based on the idea that errors follow a
normal distribution. The resulting graphs from a GENOID adjustment illustrate the directions of expansion within the parameter space. These graphs are plotted with an adjusted parameter of the dynamic
model on the x-axis and the fitness function (labelled as "fp" in the figures), in our case the $\chi^2$, on the y-axis.

These graphs can present three distinct shapes: a U-shaped curve, a significant slope, or oscillations with local minima. The U-shaped curve indicates that the solution is well-converged. Therefore, small modifications to the initial conditions on either side of the optimal solution result in a degradation of the fit. In the case of oscillations with multiple local minima, it is necessary to explore each possibility by narrowing the search space around each minimum and recalculating to analyse how each restricted system behaves. Generally, the absolute minimum is chosen as the final solution. Finally, when the graphs show a slope, this suggests that the search space must be shifted in the direction that minimizes the $\chi^2$. The calculations are then repeated until a solution is
obtained that produces a U-shaped graph.

\begin{figure}
    \centering
    \includegraphics[width=0.45\linewidth]{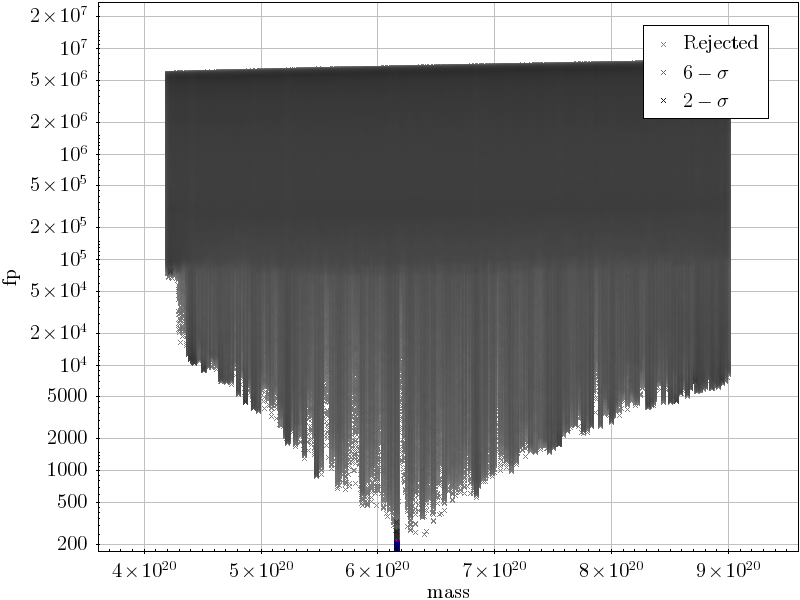}
    \includegraphics[width=0.45\linewidth]{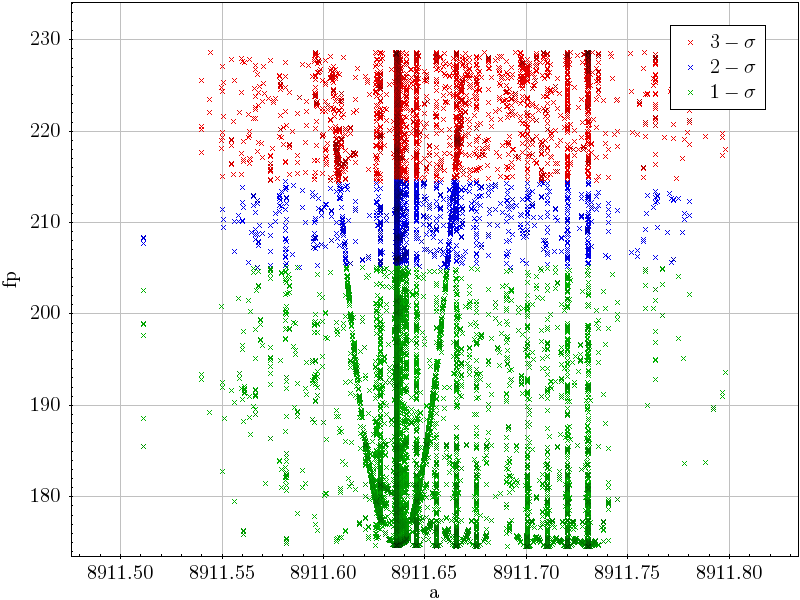}
    \includegraphics[width=0.45\linewidth]{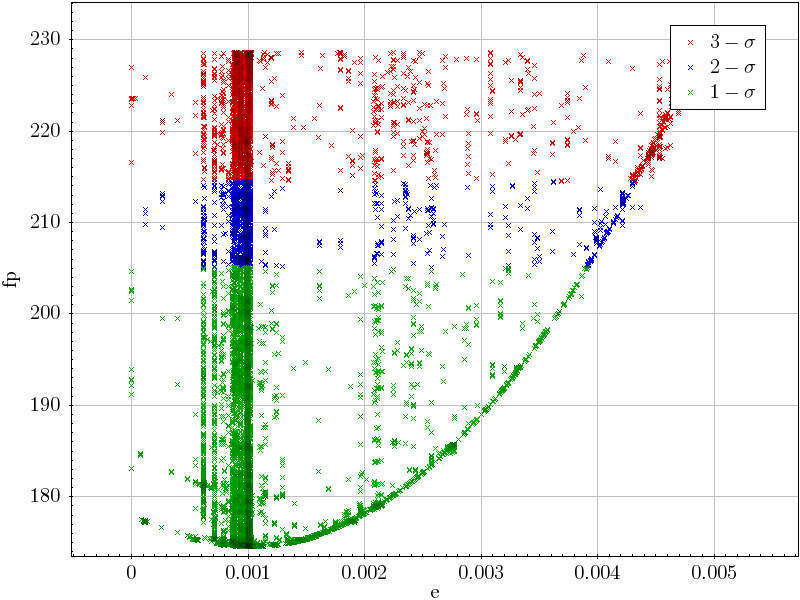}
    \includegraphics[width=0.45\linewidth]{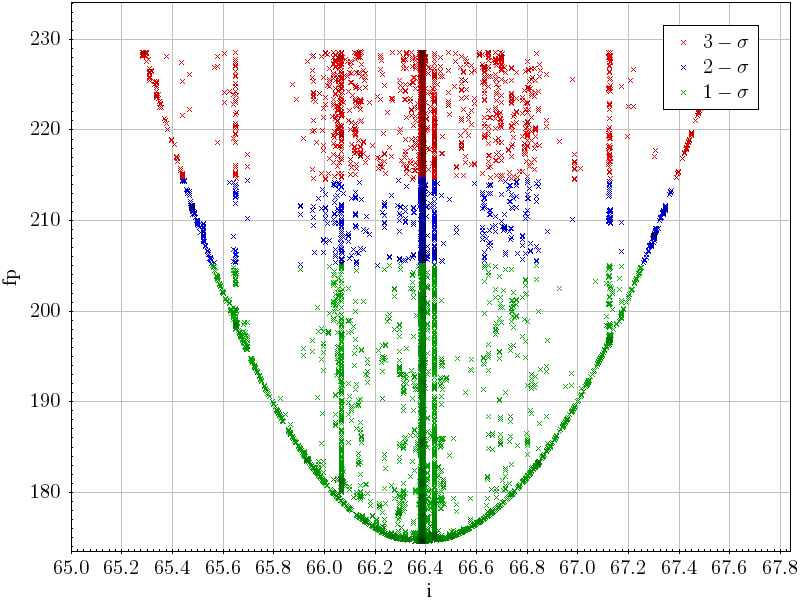}
    \caption{ The $\chi^2$ values are plotted versus some of the fitted parameters for the prograde solution of Vanth's orbit around Orcus. The top left panel presents all the space of parameters explored for the fit of the system mass, showing a well-defined minimum. The top right panel shows the 1 to 3$\sigma$ regions for the semi-major axis $a$. The bottom left and right panels show the same for the orbital eccentricity $e$ and inclination $i$, respectively. The eccentricity for the prograde solution has its minimum close to, but not at zero, although the circular solution cannot be discarded.}
    \label{fig:chiVanthPrograde}
\end{figure}

\begin{figure}
    \centering
    \includegraphics[width=0.45\linewidth]{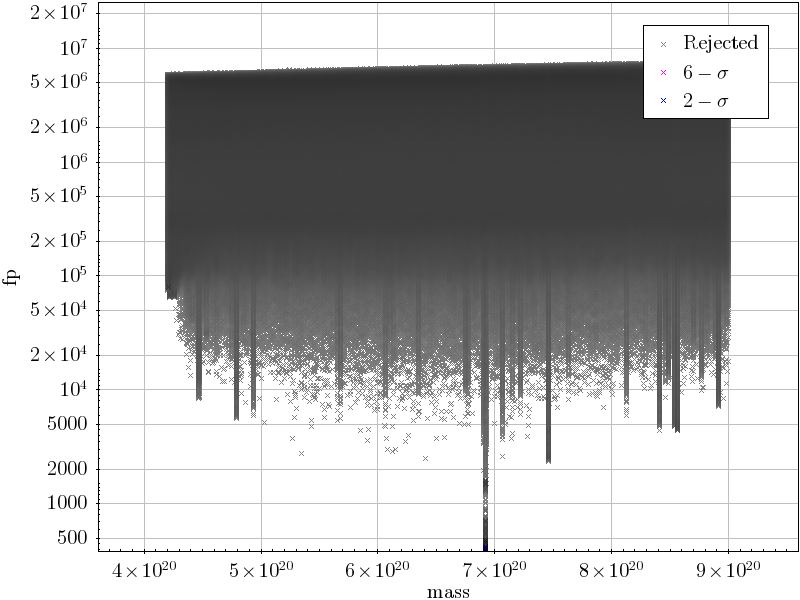}
    \includegraphics[width=0.45\linewidth]{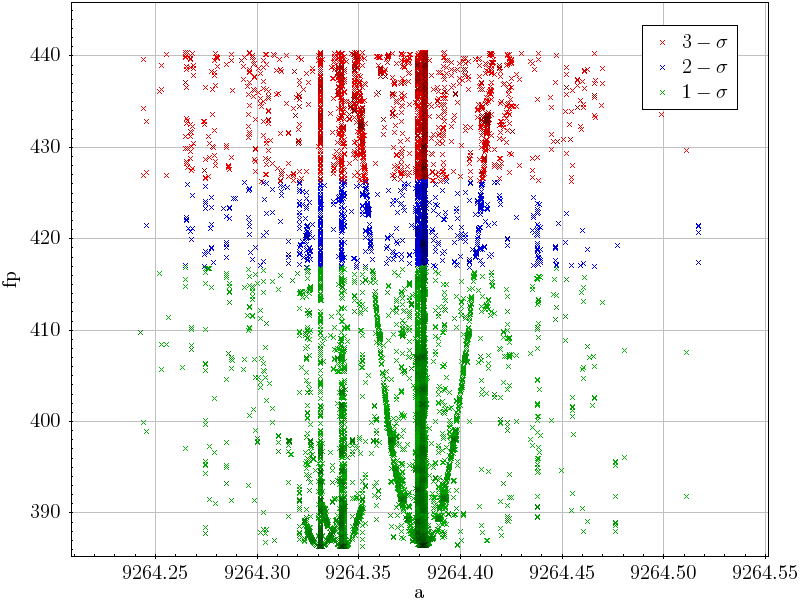}
    \includegraphics[width=0.45\linewidth]{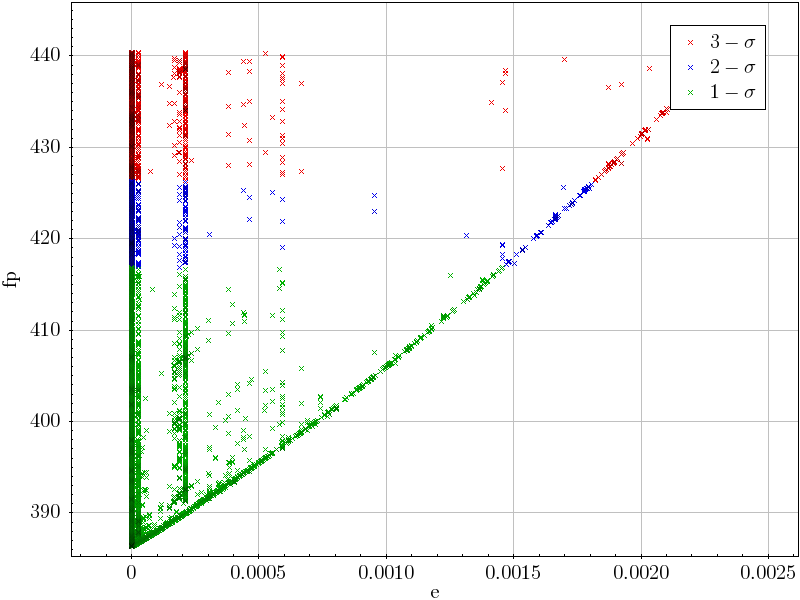}
    \includegraphics[width=0.45\linewidth]{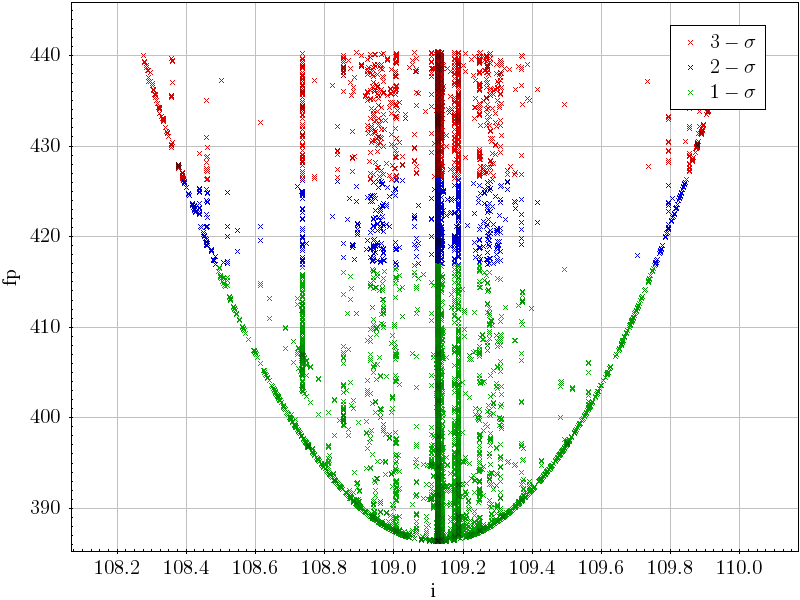}
    \caption{ The same as Fig.~\ref{fig:chiVanthPrograde} for the retrograde solution of Vanth's orbit. Again, the top left panel displays a well-defined minimum. The bottom left panel shows that the best retrograde solution is the circular orbit.  }
    \label{fig:chiVanthRetro}
\end{figure}

\begin{figure}
    \centering
    \includegraphics[width=0.45\linewidth]{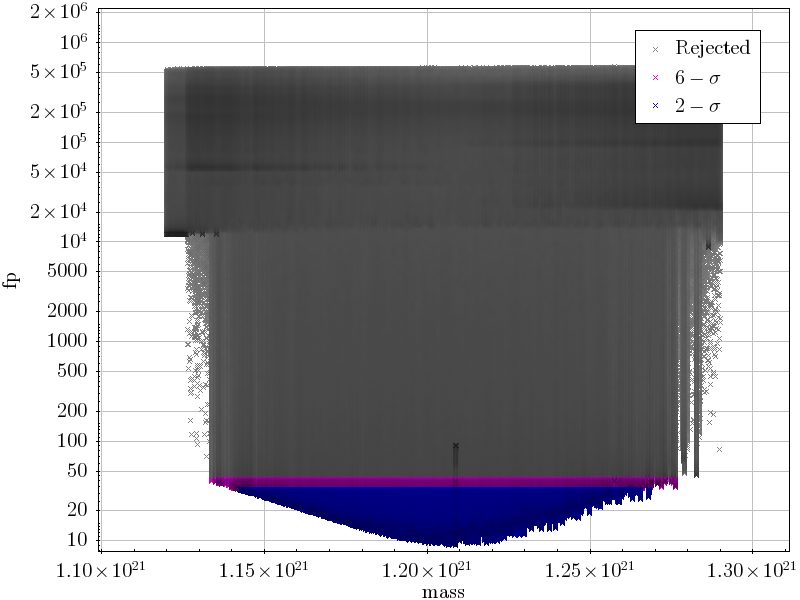}
    \includegraphics[width=0.45\linewidth]{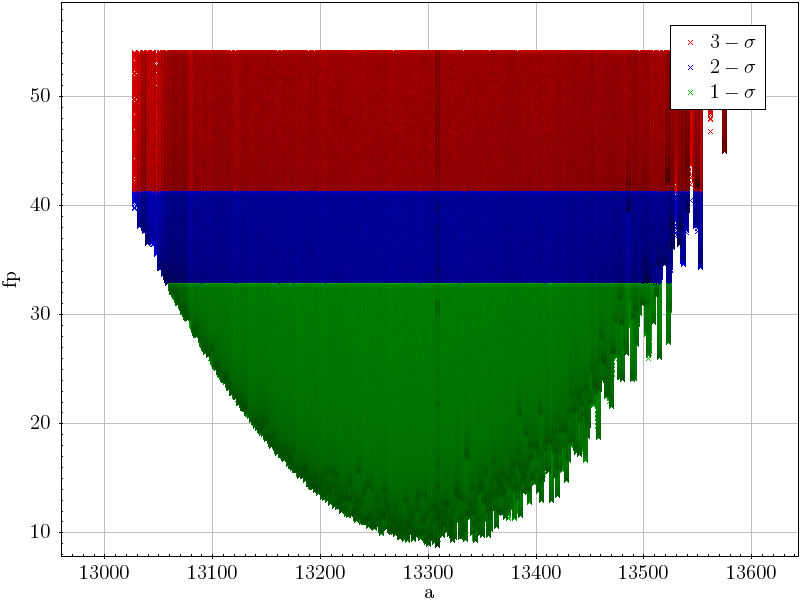}
    \includegraphics[width=0.45\linewidth]{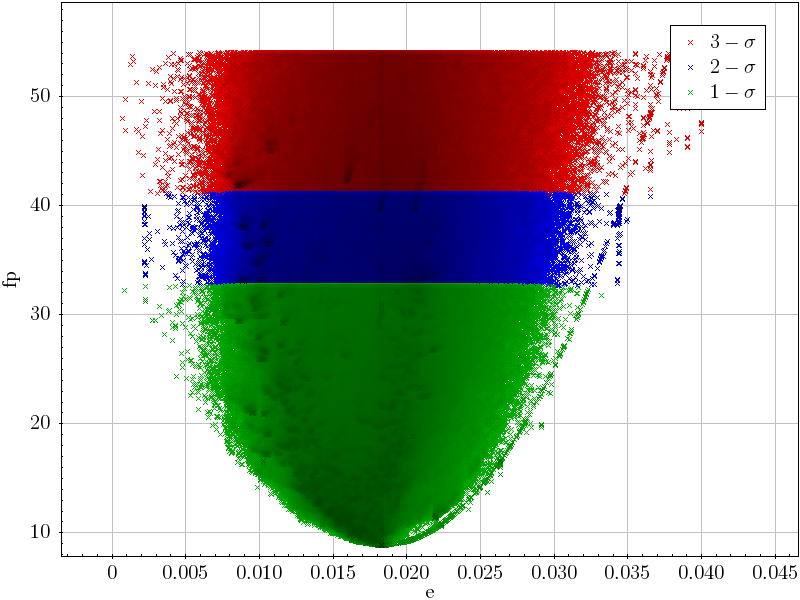}
    \includegraphics[width=0.45\linewidth]{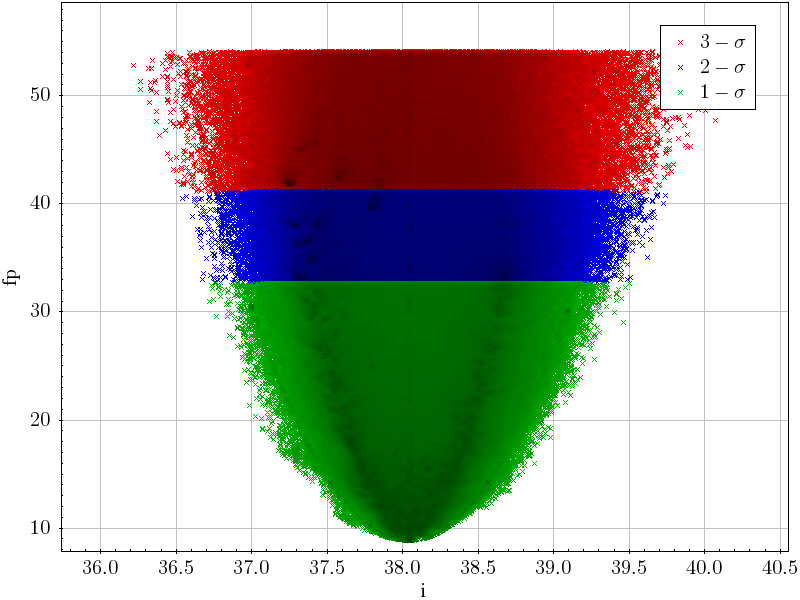}
    \caption{ The same as Fig.~\ref{fig:chiVanthPrograde} for the Weywot orbit solution around Quaoar.}
    \label{fig:chiWeywot}
\end{figure}

Figure~\ref{fig:chiVanthPrograde} shows the curves for Vanth (Orcus/1) prograde solution, while Fig.~\ref{fig:chiVanthRetro} shows the curves for Vanth (Orcus/1) retrograde solution and Fig.~\ref{fig:chiWeywot} shows the curves for Weywot (Quaoar/1) orbital solution.

\end{appendices}

% F I M.

%%%%%%%%%% Insert bibliography here %%%%%%%%%%%%%%

\bibliographystyle{RS}
\bibliography{biblio}

\begin{thebibliography}{99}

\bibitem{Levison03}
{Levison} HF, {Morbidelli} A. 2003  {The formation of the Kuiper belt by the outward transport of bodies during Neptune's migration}. {\em \nat} \textbf{426}, 419--421.
(\href{http://dx.doi.org/10.1038/nature02120}{10.1038/nature02120})

\bibitem{Gomes04}
{Gomes} RS, {Morbidelli} A, {Levison} HF. 2004  {Planetary migration in a planetesimal disk: why did Neptune stop at 30 AU?}. {\em \icarus} \textbf{170}, 492--507.
(\href{http://dx.doi.org/10.1016/j.icarus.2004.03.011}{10.1016/j.icarus.2004.03.011})

\bibitem{Tsiganis05}
{Tsiganis} K, {Gomes} R, {Morbidelli} A, {Levison} HF. 2005  {Origin of the orbital architecture of the giant planets of the Solar System}. {\em \nat} \textbf{435}, 459--461.
(\href{http://dx.doi.org/10.1038/nature03539}{10.1038/nature03539})

\bibitem{Nesvorny16}
{Nesvorn{\'y}} D, {Vokrouhlick{\'y}} D. 2016  {Neptune's Orbital Migration Was Grainy, Not Smooth}. {\em \apj} \textbf{825}, 94.
(\href{http://dx.doi.org/10.3847/0004-637X/825/2/94}{10.3847/0004-637X/825/2/94})

\bibitem{Gladman08}
{Gladman} B, {Marsden} BG, {Vanlaerhoven} C. 2008  {Nomenclature in the Outer Solar System}. In {Barucci} MA, {Boehnhardt} H, {Cruikshank} DP, {Morbidelli} A, {Dotson} R, editors, {\em The Solar System Beyond Neptune} ,  pp. 43--57. University of Arizona Press.

\bibitem{Gladman21}
{Gladman} B, {Volk} K. 2021  {Transneptunian Space}. {\em \araa} \textbf{59}, 203--246.
(\href{http://dx.doi.org/10.1146/annurev-astro-120920-010005}{10.1146/annurev-astro-120920-010005})

\bibitem{ParkerKavelaars10}
{Parker} AH, {Kavelaars} JJ. 2010  {Destruction of Binary Minor Planets During Neptune Scattering}. {\em \apjl} \textbf{722}, L204--L208.
(\href{http://dx.doi.org/10.1088/2041-8205/722/2/L204}{10.1088/2041-8205/722/2/L204})

\bibitem{ThirouinSheppard2019}
{Thirouin} A, {Sheppard} SS. 2019  {Light Curves and Rotational Properties of the Pristine Cold Classical Kuiper Belt Objects}. {\em \aj} \textbf{157}, 228.
(\href{http://dx.doi.org/10.3847/1538-3881/ab18a9}{10.3847/1538-3881/ab18a9})

\bibitem{Sicardy2024}
Sicardy B, Braga-Ribas F, Buie MW, Ortiz JL, Roques F. 2024  {Stellar occultations by trans-Neptunian objects}. {\em The Astronomy and Astrophysics Review} \textbf{32}, 6.
(\href{http://dx.doi.org/10.1007/s00159-024-00156-x}{10.1007/s00159-024-00156-x})

\bibitem{Ortiz20}
{Ortiz} JL, {Sicardy} B, {Camargo} JIB, {Santos-Sanz} P, {Braga-Ribas} F. 2020  {Stellar occultation by TNOs: from predictions to observations}. In {Prialnik} D, {Barucci} MA, {Young} L, editors, {\em The Trans-Neptunian Solar System} ,  pp. 413--437. Elsevier.
(\href{http://dx.doi.org/10.1016/B978-0-12-816490-7.00019-9}{10.1016/B978-0-12-816490-7.00019-9})

\bibitem{Braga-Ribas2013}
{Braga-Ribas} F, {Sicardy} B, {Ortiz} JL, {Lellouch} E, {Tancredi} G, {Lecacheux} J, {Vieira-Martins} R, {Camargo} JIB, {Assafin} M, {Behrend} R, {Vachier} F, {Colas} F, {Morales} N, {Maury} A, {Emilio} M, {Amorim} A, {Unda-Sanzana} E, {Roland} S, {Bruzzone} S, {Almeida} LA, {Rodrigues} CV, {Jacques} C, {Gil-Hutton} R, {Vanzi} L, {Milone} AC, {Schoenell} W, {Salvo} R, {Almenares} L, {Jehin} E, {Manfroid} J, {Sposetti} S, {Tanga} P, {Klotz} A, {Frappa} E, {Cacella} P, {Colque} JP, {Neves} C, {Alvarez} EM, {Gillon} M, {Pimentel} E, {Giacchini} B, {Roques} F, {Widemann} T, {Magalh{\~a}es} VS, {Thirouin} A, {Duffard} R, {Leiva} R, {Toledo} I, {Capeche} J, {Beisker} W, {Pollock} J, {Cede{\~n}o Monta{\~n}a} CE, {Ivarsen} K, {Reichart} D, {Haislip} J, {Lacluyze} A. 2013  {The Size, Shape, Albedo, Density, and Atmospheric Limit of Transneptunian Object (50000) Quaoar from Multi-chord Stellar Occultations}. {\em \apj} \textbf{773}, 26.
(\href{http://dx.doi.org/10.1088/0004-637X/773/1/26}{10.1088/0004-637X/773/1/26})

\bibitem{Ortiz2017}
{Ortiz} JL, {Santos-Sanz} P, {Sicardy} B, {Benedetti-Rossi} G, {B{\'e}rard} D, {Morales} N, {Duffard} R, {Braga-Ribas} F, {Hopp} U, {Ries} C, {Nascimbeni} V, {Marzari} F, {Granata} V, {P{\'a}l} A, {Kiss} C, {Pribulla} T, {Kom{\v{z}}{\'\i}k} R, {Hornoch} K, {Pravec} P, {Bacci} P, {Maestripieri} M, {Nerli} L, {Mazzei} L, {Bachini} M, {Martinelli} F, {Succi} G, {Ciabattari} F, {Mikuz} H, {Carbognani} A, {Gaehrken} B, {Mottola} S, {Hellmich} S, {Rommel} FL, {Fern{\'a}ndez-Valenzuela} E, {Campo Bagatin} A, {Cikota} S, {Cikota} A, {Lecacheux} J, {Vieira-Martins} R, {Camargo} JIB, {Assafin} M, {Colas} F, {Behrend} R, {Desmars} J, {Meza} E, {Alvarez-Candal} A, {Beisker} W, {Gomes-Junior} AR, {Morgado} BE, {Roques} F, {Vachier} F, {Berthier} J, {Mueller} TG, {Madiedo} JM, {Unsalan} O, {Sonbas} E, {Karaman} N, {Erece} O, {Koseoglu} DT, {Ozisik} T, {Kalkan} S, {Guney} Y, {Niaei} MS, {Satir} O, {Yesilyaprak} C, {Puskullu} C, {Kabas} A, {Demircan} O, {Alikakos} J, {Charmandaris} V, {Leto} G, {Ohlert} J, {Christille} JM,
  {Szak{\'a}ts} R, {Tak{\'a}csn{\'e} Farkas} A, {Varga-Vereb{\'e}lyi} E, {Marton} G, {Marciniak} A, {Bartczak} P, {Santana-Ros} T, {Butkiewicz-B{\k{a}}k} M, {Dudzi{\'n}ski} G, {Al{\'\i}-Lagoa} V, {Gazeas} K, {Tzouganatos} L, {Paschalis} N, {Tsamis} V, {S{\'a}nchez-Lavega} A, {P{\'e}rez-Hoyos} S, {Hueso} R, {Guirado} JC, {Peris} V, {Iglesias-Marzoa} R. 2017  {The size, shape, density and ring of the dwarf planet Haumea from a stellar occultation}. {\em \nat} \textbf{550}, 219--223.
(\href{http://dx.doi.org/10.1038/nature24051}{10.1038/nature24051})

\bibitem{Morgado2023}
{Morgado} BE, {Sicardy} B, {Braga-Ribas} F, {Ortiz} JL, {Salo} H, {Vachier} F, {Desmars} J, {Pereira} CL, {Santos-Sanz} P, {Sfair} R, {de Santana} T, {Assafin} M, {Vieira-Martins} R, {Gomes-J{\'u}nior} AR, {Margoti} G, {Dhillon} VS, {Fern{\'a}ndez-Valenzuela} E, {Broughton} J, {Bradshaw} J, {Langersek} R, {Benedetti-Rossi} G, {Souami} D, {Holler} BJ, {Kretlow} M, {Boufleur} RC, {Camargo} JIB, {Duffard} R, {Beisker} W, {Morales} N, {Lecacheux} J, {Rommel} FL, {Herald} D, {Benz} W, {Jehin} E, {Jankowsky} F, {Marsh} TR, {Littlefair} SP, {Bruno} G, {Pagano} I, {Brandeker} A, {Collier-Cameron} A, {Flor{\'e}n} HG, {Hara} N, {Olofsson} G, {Wilson} TG, {Benkhaldoun} Z, {Busuttil} R, {Burdanov} A, {Ferrais} M, {Gault} D, {Gillon} M, {Hanna} W, {Kerr} S, {Kolb} U, {Nosworthy} P, {Sebastian} D, {Snodgrass} C, {Teng} JP, {de Wit} J. 2023  {A dense ring of the trans-Neptunian object Quaoar outside its Roche limit}. {\em \nat} \textbf{614}, 239--243.
(\href{http://dx.doi.org/10.1038/s41586-022-05629-6}{10.1038/s41586-022-05629-6})

\bibitem{Pereira2023}
{Pereira} CL, {Sicardy} B, {Morgado} BE, {Braga-Ribas} F, {Fern{\'a}ndez-Valenzuela} E, {Souami} D, {Holler} BJ, {Boufleur} RC, {Margoti} G, {Assafin} M, {Ortiz} JL, {Santos-Sanz} P, {Epinat} B, {Kervella} P, {Desmars} J, {Vieira-Martins} R, {Kilic} Y, {Gomes J{\'u}nior} AR, {Camargo} JIB, {Emilio} M, {Vara-Lubiano} M, {Kretlow} M, {Albert} L, {Alcock} C, {Ball} JG, {Bender} K, {Buie} MW, {Butterfield} K, {Camarca} M, {Castro-Chac{\'o}n} JH, {Dunford} R, {Fisher} RS, {Gamble} D, {Geary} JC, {Gnilka} CL, {Green} KD, {Hartman} ZD, {Huang} CK, {Januszewski} H, {Johnston} J, {Kagitani} M, {Kamin} R, {Kavelaars} JJ, {Keller} JM, {de Kleer} KR, {Lehner} MJ, {Luken} A, {Marchis} F, {Marlin} T, {McGregor} K, {Nikitin} V, {Nolthenius} R, {Patrick} C, {Redfield} S, {Rengstorf} AW, {Reyes-Ruiz} M, {Seccull} T, {Skrutskie} MF, {Smith} AB, {Sproul} M, {Stephens} AW, {Szentgyorgyi} A, {S{\'a}nchez-Sanju{\'a}n} S, {Tatsumi} E, {Verbiscer} A, {Wang} SY, {Yoshida} F, {Young} R, {Zhang} ZW. 2023  {The two rings of (50000)
  Quaoar}. {\em \aap} \textbf{673}, L4.
(\href{http://dx.doi.org/10.1051/0004-6361/202346365}{10.1051/0004-6361/202346365})

\bibitem{Meza19}
{Meza} E, {Sicardy} B, {Assafin} M, {Ortiz} JL, {Bertrand} T, {Lellouch} E, {Desmars} J, {Forget} F, {B{\'e}rard} D, {Doressoundiram} A, {Lecacheux} J, {Oliveira} JM, {Roques} F, {Widemann} T, {Colas} F, {Vachier} F, {Renner} S, {Leiva} R, {Braga-Ribas} F, {Benedetti-Rossi} G, {Camargo} JIB, {Dias-Oliveira} A, {Morgado} B, {Gomes-J{\'u}nior} AR, {Vieira-Martins} R, {Behrend} R, {Tirado} AC, {Duffard} R, {Morales} N, {Santos-Sanz} P, {Jel{\'\i}nek} M, {Cunniffe} R, {Querel} R, {Harnisch} M, {Jansen} R, {Pennell} A, {Todd} S, {Ivanov} VD, {Opitom} C, {Gillon} M, {Jehin} E, {Manfroid} J, {Pollock} J, {Reichart} DE, {Haislip} JB, {Ivarsen} KM, {LaCluyze} AP, {Maury} A, {Gil-Hutton} R, {Dhillon} V, {Littlefair} S, {Marsh} T, {Veillet} C, {Bath} KL, {Beisker} W, {Bode} HJ, {Kretlow} M, {Herald} D, {Gault} D, {Kerr} S, {Pavlov} H, {Farag{\'o}} O, {Kl{\"o}s} O, {Frappa} E, {Lavayssi{\`e}re} M, {Cole} AA, {Giles} AB, {Greenhill} JG, {Hill} KM, {Buie} MW, {Olkin} CB, {Young} EF, {Young} LA, {Wasserman} LH,
  {Devog{\`e}le} M, {French} RG, {Bianco} FB, {Marchis} F, {Brosch} N, {Kaspi} S, {Polishook} D, {Manulis} I, {Ait Moulay Larbi} M, {Benkhaldoun} Z, {Daassou} A, {El Azhari} Y, {Moulane} Y, {Broughton} J, {Milner} J, {Dobosz} T, {Bolt} G, {Lade} B, {Gilmore} A, {Kilmartin} P, {Allen} WH, {Graham} PB, {Loader} B, {McKay} G, {Talbot} J, {Parker} S, {Abe} L, {Bendjoya} P, {Rivet} JP, {Vernet} D, {Di Fabrizio} L, {Lorenzi} V, {Magazz{\'u}} A, {Molinari} E, {Gazeas} K, {Tzouganatos} L, {Carbognani} A, {Bonnoli} G, {Marchini} A, {Leto} G, {Sanchez} RZ, {Mancini} L, {Kattentidt} B, {Dohrmann} M, {Guhl} K, {Rothe} W, {Walzel} K, {Wortmann} G, {Eberle} A, {Hampf} D, {Ohlert} J, {Krannich} G, {Murawsky} G, {G{\"a}hrken} B, {Gloistein} D, {Alonso} S, {Rom{\'a}n} A, {Communal} JE, {Jabet} F, {deVisscher} S, {S{\'e}rot} J, {Janik} T, {Moravec} Z, {Machado} P, {Selva} A, {Perell{\'o}} C, {Rovira} J, {Conti} M, {Papini} R, {Salvaggio} F, {Noschese} A, {Tsamis} V, {Tigani} K, {Barroy} P, {Irzyk} M, {Neel} D, {Godard} JP,
  {Lanoisel{\'e}e} D, {Sogorb} P, {V{\'e}rilhac} D, {Bretton} M, {Signoret} F, {Ciabattari} F, {Naves} R, {Boutet} M, {De Queiroz} J, {Lindner} P, {Lindner} K, {Enskonatus} P, {Dangl} G, {Tordai} T, {Eichler} H, {Hattenbach} J, {Peterson} C, {Molnar} LA, {Howell} RR. 2019  {Lower atmosphere and pressure evolution on Pluto from ground-based stellar occultations, 1988-2016}. {\em \aap} \textbf{625}, A42.
(\href{http://dx.doi.org/10.1051/0004-6361/201834281}{10.1051/0004-6361/201834281})

\bibitem{DiasOliveira17}
{Dias-Oliveira} A, {Sicardy} B, {Ortiz} JL, {Braga-Ribas} F, {Leiva} R, {Vieira-Martins} R, {Benedetti-Rossi} G, {Camargo} JIB, {Assafin} M, {Gomes-J{\'u}nior} AR, {Baug} T, {Chandrasekhar} T, {Desmars} J, {Duffard} R, {Santos-Sanz} P, {Ergang} Z, {Ganesh} S, {Ikari} Y, {Irawati} P, {Jain} J, {Liying} Z, {Richichi} A, {Shengbang} Q, {Behrend} R, {Benkhaldoun} Z, {Brosch} N, {Daassou} A, {Frappa} E, {Gal-Yam} A, {Garcia-Lozano} R, {Gillon} M, {Jehin} E, {Kaspi} S, {Klotz} A, {Lecacheux} J, {Mahasena} P, {Manfroid} J, {Manulis} I, {Maury} A, {Mohan} V, {Morales} N, {Ofek} E, {Rinner} C, {Sharma} A, {Sposetti} S, {Tanga} P, {Thirouin} A, {Vachier} F, {Widemann} T, {Asai} A, {Hayato} W, {Hiroyuki} W, {Owada} M, {Yamamura} H, {Hayamizu} T, {Bradshaw} J, {Kerr} S, {Tomioka} H, {Andersson} S, {Dangl} G, {Haymes} T, {Naves} R, {Wortmann} G. 2017  {Study of the Plutino Object (208996) 2003 AZ$_{84}$ from Stellar Occultations: Size, Shape, and Topographic Features}. {\em \aj} \textbf{154}, 22.
(\href{http://dx.doi.org/10.3847/1538-3881/aa74e9}{10.3847/1538-3881/aa74e9})

\bibitem{Rommel2023}
{Rommel} FL, {Braga-Ribas} F, {Ortiz} JL, {Sicardy} B, {Santos-Sanz} P, {Desmars} J, {Camargo} JIB, {Vieira-Martins} R, {Assafin} M, {Morgado} BE, {Boufleur} RC, {Benedetti-Rossi} G, {Gomes-J{\'u}nior} AR, {Fern{\'a}ndez-Valenzuela} E, {Holler} BJ, {Souami} D, {Duffard} R, {Margoti} G, {Vara-Lubiano} M, {Lecacheux} J, {Plouvier} JL, {Morales} N, {Maury} A, {Fabrega} J, {Ceravolo} P, {Jehin} E, {Albanese} D, {Mariey} H, {Cikota} S, {Ru{\v{z}}djak} D, {Cikota} A, {Szak{\'a}ts} R, {Baba Aissa} D, {Gringahcene} Z, {Kashuba} V, {Koshkin} N, {Zhukov} V, {Fi{\c{s}}ek} S, {{\c{C}}akir} O, {{\"O}zer} S, {Schnabel} C, {Schnabel} M, {Signoret} F, {Morrone} L, {Santana-Ros} T, {Pereira} CL, {Emilio} M, {Burdanov} AY, {de Wit} J, {Barkaoui} K, {Gillon} M, {Leto} G, {Frasca} A, {Catanzaro} G, {Sanchez} RZ, {Tagliaferri} U, {Di Sora} M, {Isopi} G, {Krugly} Y, {Slyusarev} I, {Chiorny} V, {Miku{\v{z}}} H, {Bacci} P, {Maestripieri} M, {Grazia} MD, {de la Cueva} I, {Yuste-Moreno} M, {Ciabattari} F, {Kozhukhov} OM,
  {Serra-Ricart} M, {Alarcon} MR, {Licandro} J, {Masi} G, {Bacci} R, {Bosch} JM, {Behem} R, {Prost} JP, {Renner} S, {Conjat} M, {Bachini} M, {Succi} G, {Stoian} L, {Juravle} A, {Carosati} D, {Gowe} B, {Carrillo} J, {Zheleznyak} AP, {Montigiani} N, {Foster} CR, {Mannucci} M, {Ruocco} N, {Cuevas} F, {Di Marcantonio} P, {Coretti} I, {Iafrate} G, {Baldini} V, {Collins} M, {P{\'a}l} A, {Cs{\'a}k} B, {Fern{\'a}ndez-Garcia} E, {Castro-Tirado} AJ, {Hudin} L, {Madiedo} JM, {Anghel} RM, {Calvo-Fern{\'a}ndez} JF, {Valvasori} A, {Guido} E, {Gherase} RM, {Kamoun} S, {Fafet} R, {S{\'a}nchez-Gonz{\'a}lez} M, {Curelaru} L, {V{\^\i}ntdevar{\u{a}}} CD, {Danescu} CA, {Gout} JF, {Schmitz} CJ, {Sota} A, {Belskaya} I, {Rodr{\'\i}guez-Marco} M, {Kilic} Y, {Frappa} E, {Klotz} A, {Lavayssi{\`e}re} M, {Oliveira} JM, {Popescu} M, {Mammana} LA, {Fern{\'a}ndez-Laj{\'u}s} E, {Schmidt} M, {Hopp} U, {Kom{\v{z}}{\'\i}k} R, {Pribulla} T, {Tomko} D, {Hus{\'a}rik} M, {Erece} O, {Eryilmaz} S, {Buzzi} L, {G{\"a}hrken} B, {Nardiello} D, {Hornoch}
  K, {Sonbas} E, {Er} H, {Burwitz} V, {Sybilski} PW, {Bykowski} W, {M{\"u}ller} TG, {Ogloza} W, {Gon{\c{c}}alves} R, {Ferreira} JF, {Ferreira} M, {Bento} M, {Meister} S, {Bagiran} MN, {Teke{\c{s}}} M, {Marciniak} A, {Moravec} Z, {Delin{\v{c}}{\'a}k} P, {Gianni} G, {Casalnuovo} GB, {Boutet} M, {Sanchez} J, {Klemt} B, {Wuensche} N, {Burzynski} W, {Borkowski} M, {Serrau} M, {Dangl} G, {Kl{\"o}s} O, {Weber} C, {Urban{\'\i}k} M, {Rousselot} L, {Kub{\'a}nek} J, {Andr{\'e}} P, {Colazo} C, {Spagnotto} J, {Sickafoose} AA, {Hueso} R, {S{\'a}nchez-Lavega} A, {Fisher} RS, {Rengstorf} AW, {Perell{\'o}} C, {Dascalu} M, {Altan} M, {Gazeas} K, {de Santana} T, {Sfair} R, {Winter} OC, {Kalkan} S, {Canales-Moreno} O, {Trigo-Rodr{\'\i}guez} JM, {Tsamis} V, {Tigani} K, {Sioulas} N, {Lekkas} G, {Bertesteanu} DN, {Dumitrescu} V, {Wilberger} AJ, {Barnes} JW, {Fieber-Beyer} SK, {Swaney} RL, {Fuentes} C, {Mendez} RA, {Dumitru} BD, {Flynn} RL, {Wake} DA. 2023  {A large topographic feature on the surface of the trans-Neptunian object
  (307261) 2002 MS$_{4}$ measured from stellar occultations}. {\em \aap} \textbf{678}, A167.
(\href{http://dx.doi.org/10.1051/0004-6361/202346892}{10.1051/0004-6361/202346892})

\bibitem{Rommel2020}
{Rommel} FL, {Braga-Ribas} F, {Desmars} J, {Camargo} JIB, {Ortiz} JL, {Sicardy} B, {Vieira-Martins} R, {Assafin} M, {Santos-Sanz} P, {Duffard} R, {Fern{\'a}ndez-Valenzuela} E, {Lecacheux} J, {Morgado} BE, {Benedetti-Rossi} G, {Gomes-J{\'u}nior} AR, {Pereira} CL, {Herald} D, {Hanna} W, {Bradshaw} J, {Morales} N, {Brimacombe} J, {Burtovoi} A, {Carruthers} T, {de Barros} JR, {Fiori} M, {Gilmore} A, {Hooper} D, {Hornoch} K, {Jacques} C, {Janik} T, {Kerr} S, {Kilmartin} P, {Winkel} JM, {Naletto} G, {Nardiello} D, {Nascimbeni} V, {Newman} J, {Ossola} A, {P{\'a}l} A, {Pimentel} E, {Pravec} P, {Sposetti} S, {Stechina} A, {Szak{\'a}ts} R, {Ueno} Y, {Zampieri} L, {Broughton} J, {Dunham} JB, {Dunham} DW, {Gault} D, {Hayamizu} T, {Hosoi} K, {Jehin} E, {Jones} R, {Kitazaki} K, {Kom{\v{z}}{\'\i}k} R, {Marciniak} A, {Maury} A, {Miku{\v{z}}} H, {Nosworthy} P, {F{\'a}brega Polleri} J, {Rahvar} S, {Sfair} R, {Siqueira} PB, {Snodgrass} C, {Sogorb} P, {Tomioka} H, {Tregloan-Reed} J, {Winter} OC. 2020  {Stellar occultations
  enable milliarcsecond astrometry for Trans-Neptunian objects and Centaurs}. {\em \aap} \textbf{644}, A40.
(\href{http://dx.doi.org/10.1051/0004-6361/202039054}{10.1051/0004-6361/202039054})

\bibitem{Fraser17}
{Fraser} WC, {Bannister} MT, {Pike} RE, {Marsset} M, {Schwamb} ME, {Kavelaars} JJ, {Lacerda} P, {Nesvorn{\'y}} D, {Volk} K, {Delsanti} A, {Benecchi} S, {Lehner} MJ, {Noll} K, {Gladman} B, {Petit} JM, {Gwyn} S, {Chen} YT, {Wang} SY, {Alexandersen} M, {Burdullis} T, {Sheppard} S, {Trujillo} C. 2017  {All planetesimals born near the Kuiper belt formed as binaries}. {\em Nature Astronomy} \textbf{1}, 0088.
(\href{http://dx.doi.org/10.1038/s41550-017-0088}{10.1038/s41550-017-0088})

\bibitem{NesvornyVok19}
{Nesvorn{\'y}} D, {Vokrouhlick{\'y}} D. 2019  {Binary survival in the outer solar system}. {\em \icarus} \textbf{331}, 49--61.
(\href{http://dx.doi.org/10.1016/j.icarus.2019.04.030}{10.1016/j.icarus.2019.04.030})

\bibitem{Levison08}
{Levison} HF, {Morbidelli} A, {Van Laerhoven} C, {Gomes} R, {Tsiganis} K. 2008  {Origin of the structure of the Kuiper belt during a dynamical instability in the orbits of Uranus and Neptune}. {\em \icarus} \textbf{196}, 258--273.
(\href{http://dx.doi.org/10.1016/j.icarus.2007.11.035}{10.1016/j.icarus.2007.11.035})

\bibitem{Leinhardt10}
{Leinhardt} ZM, {Marcus} RA, {Stewart} ST. 2010  {The Formation of the Collisional Family Around the Dwarf Planet Haumea}. {\em \apj} \textbf{714}, 1789--1799.
(\href{http://dx.doi.org/10.1088/0004-637X/714/2/1789}{10.1088/0004-637X/714/2/1789})

\bibitem{ThirouinSheppard2018}
{Thirouin} A, {Sheppard} SS. 2018  {The Plutino Population: An Abundance of Contact Binaries}. {\em \aj} \textbf{155}, 248.
(\href{http://dx.doi.org/10.3847/1538-3881/aac0ff}{10.3847/1538-3881/aac0ff})

\bibitem{Grundy19}
{Grundy} WM, {Noll} KS, {Roe} HG, {Buie} MW, {Porter} SB, {Parker} AH, {Nesvorn{\'y}} D, {Levison} HF, {Benecchi} SD, {Stephens} DC, {Trujillo} CA. 2019  {Mutual orbit orientations of transneptunian binaries}. {\em \icarus} \textbf{334}, 62--78.
(\href{http://dx.doi.org/10.1016/j.icarus.2019.03.035}{10.1016/j.icarus.2019.03.035})

\bibitem{Porter24}
{Porter} SB, {Benecchi} SD, {Verbiscer} AJ, {Grundy} WM, {Noll} KS, {Parker} AH. 2024  {Detection of Close Kuiper Belt Binaries with HST WFC3}. {\em \psj} \textbf{5}, 143.
(\href{http://dx.doi.org/10.3847/PSJ/ad3f19}{10.3847/PSJ/ad3f19})

\bibitem{Sicardy11}
{Sicardy} B, {Bolt} G, {Broughton} J, {Dobosz} T, {Gault} D, {Kerr} S, {B{\'e}nard} F, {Frappa} E, {Lecacheux} J, {Peyrot} A, {Teng-Chuen-Yu} JP, {Beisker} W, {Boissel} Y, {Buckley} D, {Colas} F, {de Witt} C, {Doressoundiram} A, {Roques} F, {Widemann} T, {Gruhn} C, {Batista} V, {Biggs} J, {Dieters} S, {Greenhill} J, {Groom} R, {Herald} D, {Lade} B, {Mathers} S, {Assafin} M, {Camargo} JIB, {Vieira-Martins} R, {Andrei} AH, {da Silva Neto} DN, {Braga-Ribas} F, {Behrend} R. 2011  {Constraints on Charon's Orbital Elements from the Double Stellar Occultation of 2008 June 22}. {\em \aj} \textbf{141}, 67.
(\href{http://dx.doi.org/10.1088/0004-6256/141/2/67}{10.1088/0004-6256/141/2/67})

\bibitem{BragaRibas17}
{Braga-Ribas} F, {Vachier} F, {Camargo} J, {Desmars} J, {Sicardy} B, {Vieira-Martins} R, {Assafin} M, {Benedetti-Rossi} G, {Dias-Oliveira} A, {Murakami} Y, {Lecacheaux} J. 2017  {Stellar Occultations by TNOs: Probing Rings, Surface, and Satellites}. In {\em Asteroids, Comets, Meteors (ACM2017), Montevideo, Uruguay}.

\bibitem{Sickafoose19}
{Sickafoose} AA, {Bosh} AS, {Levine} SE, {Zuluaga} CA, {Genade} A, {Schindler} K, {Lister} TA, {Person} MJ. 2019  {A stellar occultation by Vanth, a satellite of (90482) Orcus}. {\em \icarus} \textbf{319}, 657--668.
(\href{http://dx.doi.org/10.1016/j.icarus.2018.10.016}{10.1016/j.icarus.2018.10.016})

\bibitem{Fernandez21}
{Fern{\'a}ndez-Valenzuela} E, {Ortiz Moreno} JL, {Holler} B, {Vara-Lubiano} M, {Morales} N, {Sicardy} B, {Vachier} F, {Desmars} J, {Braga-Ribas} F, {Jehin} E, {Rustamkulov} Z, {de la Vega} A, {Warner} E, {Benkhaldoun} Z, {Kamin} R, {Ryan} A, {Earls} B, {Conti} D, {de Wit} J, {Burdanov} A, {Richard} F, {Langill} P, {Morales} R, {Fraser} W, {Souami} D, {Lecacheux} J, {Santos-Sanz} P, {Duffard} R, {Alvarez-Candal} A, {Kretlow} M, {Benedetti-Rossi} G, {Morgado} B, {Camargo} J, {Rommel} FL, {Ramos Gomes Junior} A, {Assafin} M, {Baba Aissa} D, {Grigahcene} Z, {Buie} M, {Licandro} J, {Alarcon} MR, {Serra-Ricart} M, {Castro-Tirado} A, {Fernandez-Garcia} EJ, {Iglesias-Marzoa} R, {Galindo} F, {P{\'e}rez} L, {Gonz{\'a}lez} H, {Canedo} P, {Blanco} O, {Gon{\c{c}}alves} R, {Rengstorf} A, {Flynn} R, {Olsen} A, {Hanna} B, {Barnes} J, {A'Hearn} JA, {Kreyche} SM, {Miller} WJ, {Mortensen} LE, {Gibson} TC, {Walker} G, {McAllister} GS, {Feiden} GA, {Froetschel} J, {Steel} S, {Encardes} D, {Fisher} RS, {Luken} A, {Holcomb} E,
  {Caton} D, {Dunford} B. 2021  {Physical properties of Hi'iaka from stellar occultation data}. In {\em AAS/Division for Planetary Sciences Meeting Abstracts} vol.~53{\em AAS/Division for Planetary Sciences Meeting Abstracts} p. 503.05.

\bibitem{Leiva20}
{Leiva} R, {Buie} MW, {Keller} JM, {Wasserman} LH, {Kavelaars} J, {Bridges} T, {Haley} SL, {Strauss} R, {Wilde} E, {Weryk} R, {Kervella} P, {Baker} R, {Bock} SA, {Conway} K, {Cota}, Juan~M. J, {Estes} JJ, {Garc{\'\i}a} ML, {Kehrli} M, {McCandless} A, {McCandless} K, {Self} E, {Settlemire} C, {Swanson} DJ, {Thompson} D, {Wise} JA. 2020  {Stellar Occultation by the Resonant Trans-Neptunian Object (523764) 2014 WC510 Reveals a Close Binary TNO}. {\em \psj} \textbf{1}, 48.
(\href{http://dx.doi.org/10.3847/PSJ/abb23d}{10.3847/PSJ/abb23d})

\bibitem{Fernandez23weywot}
{Fernandez-Valenzuela} E, {Holler} B, {Ortiz} JL, {Vachier} F, {Braga Ribas} F, {Rommel} F, {Desmars} J, {Buie} M, {Levine} S, {Collins} M, {Nikitin} V, {Skrutskie} M, {Collyer} C, {Pike} R, {Leiva} R, {Margoti} G, {Morales} N, {Verbiscer} A, {Stansberry} J, {DeColibus} D, {Castro Chac{\'o}n} J, {Chanover} N, {McMillan} R, {Boyle} R, {Golden} A, {Butler} R, {Ryan} M, {Strauss} R, {Zigo} H, {Porter} S, {Kao} M, {Kretlow} M, {Vara-Lubiano} M, {Sicardy} B, {Pereira} C, {Santos-Sanz} P, {Morgado} B, {Benedetti-Rossi} G. 2023  {Weywot: the darkest known satellite in the trans-Neptunian region}. In {\em AAS/Division for Planetary Sciences Meeting Abstracts} vol.~55{\em AAS/Division for Planetary Sciences Meeting Abstracts} p. 202.04.

\bibitem{Rommel25}
{Rommel} FL, {Fern{\'a}ndez-Valenzuela} E, {Proudfoot} BCN, {Ortiz} JL, {Morgado} BE, {Sicardy} B, {Morales} N, {Braga-Ribas} F, {Desmars} J, {Vieira-Martins} R, {Holler} BJ, {Kilic} Y, {Grundy} W, {Rizos} JL, {Camargo} JIB, {Benedetti-Rossi} G, {Gomes-J{\'u}nior} A, {Assafin} M, {Santos-Sanz} P, {Kretlow} M, {Vara-Lubiano} M, {Leiva} R, {Ragozzine} DA, {Duffard} R, {Ku{\v{c}}{\'a}kov{\'a}} H, {Hornoch} K, {Nikitin} V, {Santana-Ros} T, {Canales-Moreno} O, {Lafuente-Aznar} D, {Calavia-Belloc} S, {Perell{\'o}} C, {Selva} A, {Organero} F, {Hernandez} LA, {de la Cueva} I, {Yuste-Moreno} M, {Garc{\'\i}a-Navarro} E, {Donate-Lucas} JE, {Izquierdo-Carri{\'o}n} L, {Iglesias-Marzoa} R, {Lacruz} E, {Gon{\c{c}}alves} R, {Staels} B, {Goossens} R, {Henden} A, {Walker} G, {Reyes} JA, {Pastor} S, {Kaspi} S, {Skrutskie} M, {Verbiscer} AJ, {Martinez} P, {Andr{\'e}} P, {Maestre} JL, {Aceituno} FJ, {Bacci} P, {Maestripieri} M, {Grazia} MD, {Castro-Tirado} AJ, {P{\'e}rez-Garcia} I, {Fern{\'a}ndez Garc{\'\i}a} EJ, {Fern{\'a}ndez}
  E, {Messner} S, {Scarfi} G, {Miku{\v{z}}} H, {Prat} J, {Martorell} P, {Nardiello} D, {Nascimbeni} V, {Sfair} R, {Siqueira} PB, {Lattari} V, {Liberato} L, {Pinheiro} TFLL, {de Santana} T, {Pereira} CL, {Alava-Amat} MA, {Ciabattari} F, {Gonz{\'a}lez-Rodriguez} H, {Schnabel} C. 2025  {Stellar occultation observations of (38628) Huya and its satellite: a detailed look into the system}. {\em arXiv e-prints} p. arXiv:2501.09739.

\bibitem{Vara23}
{Vara-Lubiano} M, {Fern{\'a}ndez-Valenzuela} E, {Kretlow} M, {Morales} N, {Benedetti-Rossi} G, {Rommel} F, {Ortiz} JL, {Sicardy} B, {Santos-Sanz} P, {Vieira-Martins} R, {Braga-Ribas} F, {Camargo} J, {Kilic} Y, {Morgado} B, {Gomes} A, {Alvarez-Candal} A, {Lecacheux} J, {Assafin} M, {Duffard} R, {Souami} D, {Desmars} J, {Mottola} S, {Sota} A, {Pal} A, {Szak{\'a}ts} R, {Kiss} C, {Kalup} C, {Derekas} A, {Zejmo} M, {Marciniak} A, {Ogloza} W, {Dangl} G, {Carbognani} A, {Stirpe} G, {Bruni} I, {Cs{\'a}nyi} I, {Skvar{\v{c}}} J, {Mikuz} H, {Meister} S, {Conjat} M, {Ciabattari} F, {Krannich} G. 2023  {Updated Size of the Trans-Neptunian Binary 2000 YW134 from a Stellar Occultation}. In {\em Planetary Sciences and Exploration of the Solar System (7th CPESS)} p. 80575.

\bibitem{Leiva23}
{Leiva} R, {Ortiz} JL, {G{\'o}mez-Lim{\'o}n} JM, {Perez} P, {Kretlow} M, {Desmars} J, {Morales} N, {Rommel} FL, {Margoti} G, {Vara-Luviano} M, {Santos-Sanz} P, {Duffard} R, {Rizos} JL, {Fernandez-Valenzuela} E, {Braga-Ribas} F, {Liberato Mendes} L, {Malacarne} M, {Gomes-Jr} AR, {Sfair} R. 2023  {(470316) 2007 OC10, (470309) 2007 JK43, and (19521) Chaos, Results from Stellar Occultations}. In {\em LPI Contributions} vol. 2851{\em LPI Contributions} p. 2527.

\bibitem{Braga-Ribas2019}
{Braga-Ribas} F, {Crispim} A, {Vieira-Martins} R, {Sicardy} B, {Ortiz} JL, {Assafin} M, {Camargo} JIB, {Desmars} J, {Lecacheux} J, {Santos-Sanz} P, {Duffard} R, {Benedetti-Rossi} G, {Gomes-J{\'u}nior} AR, {Morgado} B, {Rommel} FL, {Margoti} G, {Pereira} CL. 2019  {Database on detected stellar occultations by small outer Solar System objects}. In {\em Journal of Physics Conference Series} vol. 1365{\em Journal of Physics Conference Series} p. 012024.
(\href{http://dx.doi.org/10.1088/1742-6596/1365/1/012024}{10.1088/1742-6596/1365/1/012024})

\bibitem{Araujo16}
{Araujo} RAN, {Sfair} R, {Winter} OC. 2016  {The Rings of Chariklo under Close Encounters with the Giant Planets}. {\em \apj} \textbf{824}, 80.
(\href{http://dx.doi.org/10.3847/0004-637X/824/2/80}{10.3847/0004-637X/824/2/80})

\bibitem{Sicardy2020}
Sicardy B, Renner S, Leiva R, Roques F, {El Moutamid} M, Santos-Sanz P, Desmars J. 2020  Chapter 11 - The dynamics of rings around Centaurs and Trans-Neptunian objects. In Prialnik D, Barucci MA, Young LA, editors, {\em The Trans-Neptunian Solar System} ,  pp. 249--269. Elsevier.
(\href{http://dx.doi.org/https://doi.org/10.1016/B978-0-12-816490-7.00011-4}{https://doi.org/10.1016/B978-0-12-816490-7.00011-4})

\bibitem{GaiaColab2023}
{Gaia Collaboration} et~al.. 2023  {Gaia Data Release 3. Summary of the content and survey properties}. {\em \aap} \textbf{674}, A1.
(\href{http://dx.doi.org/10.1051/0004-6361/202243940}{10.1051/0004-6361/202243940})

\bibitem{Vachier12}
{Vachier} F, {Berthier} J, {Marchis} F. 2012  {Determination of binary asteroid orbits with a genetic-based algorithm}. {\em \aap} \textbf{543}, A68.
(\href{http://dx.doi.org/10.1051/0004-6361/201118408}{10.1051/0004-6361/201118408})

\bibitem{Vachier22}
{Vachier} F, {Carry} B, {Berthier} J. 2022  {Dynamics of the binary asteroid (379) Huenna}. {\em \icarus} \textbf{382}, 115013.
(\href{http://dx.doi.org/10.1016/j.icarus.2022.115013}{10.1016/j.icarus.2022.115013})

\bibitem{Yang16}
{Yang} B, {Wahhaj} Z, {Beauvalet} L, {Marchis} F, {Dumas} C, {Marsset} M, {Nielsen} EL, {Vachier} F. 2016  {Extreme AO Observations of Two Triple Asteroid Systems with SPHERE}. {\em \apjl} \textbf{820}, L35.
(\href{http://dx.doi.org/10.3847/2041-8205/820/2/L35}{10.3847/2041-8205/820/2/L35})

\bibitem{Proudfoot24}
{Proudfoot} BCN, {Ragozzine} DA, {Thatcher} ML, {Grundy} W, {Spencer} DJ, {Alailima} TM, {Allen} S, {Bowden} PC, {Byrd} S, {Camacho} CD, {Campbell} GH, {Carlisle} EP, {Christensen} JA, {Christensen} NK, {Clement} K, {Derieg} BJ, {Dille} MK, {Dorrett} C, {Ellefson} AL, {Fleming} TS, {Freeman} NJ, {Gibson} EJ, {Giforos} WG, {Guerrette} JA, {Haddock} O, {Hammond} SA, {Hampson} ZA, {Hancock} JD, {Harmer} MS, {Henderson} JR, {Jensen} CR, {Jensen} D, {Jensen} RE, {Jones} JS, {Kubal} CC, {Lunt} JN, {Martins} S, {Matheson} M, {Maxwell} D, {Morrell} TD, {Myckowiak} MM, {Nelsen} MA, {Neu} ST, {Nuccitelli} GG, {Reardon} KM, {Reid} AS, {Richards} KG, {Robertson} MRW, {Rydalch} TD, {Scoresby} CB, {Scott} RL, {Shakespear} ZD, {Silveira} EA, {Steed} GC, {Suggs} CZ, {Suggs} GD, {Tobias} DM, {Toole} ML, {Townsend} ML, {Vickers} KL, {Wagner} CR, {Wright} MS, {Zappala} EMA. 2024  {Beyond Point Masses. II. Non-Keplerian Shape Effects Are Detectable in Several TNO Binaries}. {\em \aj} \textbf{167}, 144.
(\href{http://dx.doi.org/10.3847/1538-3881/ad26f0}{10.3847/1538-3881/ad26f0})

\bibitem{desmars2015}
{Desmars} J et~al.. 2015  {Orbit determination of trans-Neptunian objects and Centaurs for the prediction of stellar occultations}. {\em \aap} \textbf{584}, A96.
(\href{http://dx.doi.org/10.1051/0004-6361/201526498}{10.1051/0004-6361/201526498})

\bibitem{Ferreira20}
{Ferreira} JF, {Tanga} P, {Machado} P, {Corsaro} E. 2020  {A survey for occultation astrometry of main belt: expected astrometric performances}. {\em \aap} \textbf{641}, A81.
(\href{http://dx.doi.org/10.1051/0004-6361/202038190}{10.1051/0004-6361/202038190})

\bibitem{Camargo18}
{Camargo} JIB, {Desmars} J, {Braga-Ribas} F, {Vieira-Martins} R, {Assafin} M, {Sicardy} B, {B{\'e}rard} D, {Benedetti-Rossi} G. 2018  {The future of stellar occultations by distant solar system bodies: Perspectives from the Gaia astrometry and the deep sky surveys}. {\em \planss} \textbf{154}, 59--62.
(\href{http://dx.doi.org/10.1016/j.pss.2018.02.014}{10.1016/j.pss.2018.02.014})

\bibitem{Rossi14}
{Benedetti-Rossi} G, {Vieira Martins} R, {Camargo} JIB, {Assafin} M, {Braga-Ribas} F. 2014  {Pluto: improved astrometry from 19 years of observations}. {\em \aap} \textbf{570}, A86.
(\href{http://dx.doi.org/10.1051/0004-6361/201424275}{10.1051/0004-6361/201424275})

\bibitem{Berdeu22}
{Berdeu} A, {Langlois} M, {Vachier} F. 2022  {First observation of a quadruple asteroid. Detection of a third moon around (130) Elektra with SPHERE/IFS}. {\em \aap} \textbf{658}, L4.
(\href{http://dx.doi.org/10.1051/0004-6361/202142623}{10.1051/0004-6361/202142623})

\bibitem{Brown23}
{Brown} ME, {Butler} BJ. 2023  {Masses and Densities of Dwarf Planet Satellites Measured with ALMA}. {\em \psj} \textbf{4}, 193.
(\href{http://dx.doi.org/10.3847/PSJ/ace52a}{10.3847/PSJ/ace52a})

\bibitem{Brown10}
{Brown} ME, {Ragozzine} D, {Stansberry} J, {Fraser} WC. 2010  {The Size, Density, and Formation of the Orcus-Vanth System in the Kuiper Belt}. {\em \aj} \textbf{139}, 2700--2705.
(\href{http://dx.doi.org/10.1088/0004-6256/139/6/2700}{10.1088/0004-6256/139/6/2700})

\bibitem{Carry11}
{Carry} B, {Hestroffer} D, {DeMeo} FE, {Thirouin} A, {Berthier} J, {Lacerda} P, {Sicardy} B, {Doressoundiram} A, {Dumas} C, {Farrelly} D, {M{\"u}ller} TG. 2011  {Integral-field spectroscopy of (90482) Orcus-Vanth}. {\em \aap} \textbf{534}, A115.
(\href{http://dx.doi.org/10.1051/0004-6361/201117486}{10.1051/0004-6361/201117486})

\bibitem{Camargo14}
{Camargo} JIB, {Vieira-Martins} R, {Assafin} M, {Braga-Ribas} F, {Sicardy} B, {Desmars} J, {Andrei} AH, {Benedetti-Rossi} G, {Dias-Oliveira} A. 2014  {Candidate stellar occultations by Centaurs and trans-Neptunian objects up to 2014}. {\em \aap} \textbf{561}, A37.
(\href{http://dx.doi.org/10.1051/0004-6361/201322579}{10.1051/0004-6361/201322579})

\bibitem{Brown18}
{Brown} ME, {Butler} BJ. 2018  {Medium-sized Satellites of Large Kuiper Belt Objects}. {\em \aj} \textbf{156}, 164.
(\href{http://dx.doi.org/10.3847/1538-3881/aad9f2}{10.3847/1538-3881/aad9f2})

\bibitem{Ortiz03}
{Ortiz} JL, {Guti{\'e}rrez} PJ, {Sota} A, {Casanova} V, {Teixeira} VR. 2003  {Rotational brightness variations in Trans-Neptunian Object 50000 Quaoar}. {\em \aap} \textbf{409}, L13--L16.
(\href{http://dx.doi.org/10.1051/0004-6361:20031253}{10.1051/0004-6361:20031253})

\bibitem{BrownSuer07}
{Brown} ME, {Suer} TA. 2007  {Satellites of 2003 AZ\_84, (50000), (55637), and (90482)}. {\em \iaucirc} \textbf{8812}, 1.

\bibitem{Fornasier13}
{Fornasier} S, {Lellouch} E, {M{\"u}ller} T, {Santos-Sanz} P, {Panuzzo} P, {Kiss} C, {Lim} T, {Mommert} M, {Bockel{\'e}e-Morvan} D, {Vilenius} E, {Stansberry} J, {Tozzi} GP, {Mottola} S, {Delsanti} A, {Crovisier} J, {Duffard} R, {Henry} F, {Lacerda} P, {Barucci} A, {Gicquel} A. 2013  {TNOs are Cool: A survey of the trans-Neptunian region. VIII. Combined Herschel PACS and SPIRE observations of nine bright targets at 70-500 {\ensuremath{\mu}}m}. {\em \aap} \textbf{555}, A15.
(\href{http://dx.doi.org/10.1051/0004-6361/201321329}{10.1051/0004-6361/201321329})

\bibitem{Fraser13}
{Fraser} WC, {Batygin} K, {Brown} ME, {Bouchez} A. 2013  {The mass, orbit, and tidal evolution of the Quaoar-Weywot system}. {\em \icarus} \textbf{222}, 357--363.
(\href{http://dx.doi.org/10.1016/j.icarus.2012.11.004}{10.1016/j.icarus.2012.11.004})

\bibitem{Elliot1989}
{Elliot} JL, {Dunham} EW, {Bosh} AS, {Slivan} SM, {Young} LA, {Wasserman} LH, {Millis} RL. 1989  {Pluto's atmosphere}. {\em \icarus} \textbf{77}, 148--170.
(\href{http://dx.doi.org/10.1016/0019-1035(89)90014-6}{10.1016/0019-1035(89)90014-6})

\bibitem{Brosch1995}
{Brosch} N. 1995  {The 1985 stellar occultation by Pluto}. {\em \mnras} \textbf{276}, 571--578.
(\href{http://dx.doi.org/10.1093/mnras/276.2.571}{10.1093/mnras/276.2.571})

\bibitem{Elliot10}
{Elliot} JL, {Person} MJ, {Zuluaga} CA, {Bosh} AS, {Adams} ER, {Brothers} TC, {Gulbis} AAS, {Levine} SE, {Lockhart} M, {Zangari} AM, {Babcock} BA, {Dupr{\'e}} K, {Pasachoff} JM, {Souza} SP, {Rosing} W, {Secrest} N, {Bright} L, {Dunham} EW, {Sheppard} SS, {Kakkala} M, {Tilleman} T, {Berger} B, {Briggs} JW, {Jacobson} G, {Valleli} P, {Volz} B, {Rapoport} S, {Hart} R, {Brucker} M, {Michel} R, {Mattingly} A, {Zambrano-Marin} L, {Meyer} AW, {Wolf} J, {Ryan} EV, {Ryan} WH, {Morzinski} K, {Grigsby} B, {Brimacombe} J, {Ragozzine} D, {Montano} HG, {Gilmore} A. 2010  {Size and albedo of Kuiper belt object 55636 from a stellar occultation}. {\em \nat} \textbf{465}, 897--900.
(\href{http://dx.doi.org/10.1038/nature09109}{10.1038/nature09109})

\bibitem{MarquesOliveira22}
{Marques Oliveira} J, {Sicardy} B, {Gomes-J{\'u}nior} AR, {Ortiz} JL, {Strobel} DF, {Bertrand} T, {Forget} F, {Lellouch} E, {Desmars} J, {B{\'e}rard} D, {Doressoundiram} A, {Lecacheux} J, {Leiva} R, {Meza} E, {Roques} F, {Souami} D, {Widemann} T, {Santos-Sanz} P, {Morales} N, {Duffard} R, {Fern{\'a}ndez-Valenzuela} E, {Castro-Tirado} AJ, {Braga-Ribas} F, {Morgado} BE, {Assafin} M, {Camargo} JIB, {Vieira-Martins} R, {Benedetti-Rossi} G, {Santos-Filho} S, {Banda-Huarca} MV, {Quispe-Huaynasi} F, {Pereira} CL, {Rommel} FL, {Margoti} G, {Dias-Oliveira} A, {Colas} F, {Berthier} J, {Renner} S, {Hueso} R, {P{\'e}rez-Hoyos} S, {S{\'a}nchez-Lavega} A, {Rojas} JF, {Beisker} W, {Kretlow} M, {Herald} D, {Gault} D, {Bath} KL, {Bode} HJ, {Bredner} E, {Guhl} K, {Haymes} TV, {Hummel} E, {Kattentidt} B, {Kl{\"o}s} O, {Pratt} A, {Thome} B, {Avdellidou} C, {Gazeas} K, {Karampotsiou} E, {Tzouganatos} L, {Kardasis} E, {Christou} AA, {Xilouris} EM, {Alikakos} I, {Gourzelas} A, {Liakos} A, {Charmandaris} V, {Jel{\'\i}nek} M,
  {{\v{S}}trobl} J, {Eberle} A, {Rapp} K, {G{\"a}hrken} B, {Klemt} B, {Kowollik} S, {Bitzer} R, {Miller} M, {Herzogenrath} G, {Frangenberg} D, {Brandis} L, {P{\"u}tz} I, {Perdelwitz} V, {Piehler} GM, {Riepe} P, {von Poschinger} K, {Baruffetti} P, {Cenadelli} D, {Christille} JM, {Ciabattari} F, {Di Luca} R, {Alboresi} D, {Leto} G, {Zanmar Sanchez} R, {Bruno} P, {Occhipinti} G, {Morrone} L, {Cupolino} L, {Noschese} A, {Vecchione} A, {Scalia} C, {Lo Savio} R, {Giardina} G, {Kamoun} S, {Barbosa} R, {Behrend} R, {Spano} M, {Bouchet} E, {Cottier} M, {Falco} L, {Gallego} S, {Tortorelli} L, {Sposetti} S, {Sussenbach} J, {Van Den Abbeel} F, {Andr{\'e}} P, {Llibre} M, {Pailler} F, {Ardissone} J, {Boutet} M, {Sanchez} J, {Bretton} M, {Cailleau} A, {Pic} V, {Granier} L, {Chauvet} R, {Conjat} M, {Dauvergne} JL, {Dechambre} O, {Delay} P, {Delcroix} M, {Rousselot} L, {Ferreira} J, {Machado} P, {Tanga} P, {Rivet} JP, {Frappa} E, {Irzyk} M, {Jabet} F, {Kaschinski} M, {Klotz} A, {Rieugnie} Y, {Klotz} AN, {Labrevoir} O,
  {Lavandier} D, {Walliang} D, {Leroy} A, {Bouley} S, {Lisciandra} S, {Coliac} JF, {Metz} F, {Erpelding} D, {Nougayr{\`e}de} P, {Midavaine} T, {Miniou} M, {Moindrot} S, {Morel} P, {Reginato} B, {Reginato} E, {Rudelle} J, {Tregon} B, {Tanguy} R, {David} J, {Thuillot} W, {Hestroffer} D, {Vaudescal} G, {Baba Aissa} D, {Grigahcene} Z, {Briggs} D, {Broadbent} S, {Denyer} P, {Haigh} NJ, {Quinn} N, {Thurston} G, {Fossey} SJ, {Arena} C, {Jennings} M, {Talbot} J, {Alonso} S, {Rom{\'a}n Reche} A, {Casanova} V, {Briggs} E, {Iglesias-Marzoa} R, {Abril Ib{\'a}{\~n}ez} J, {D{\'\i}az Mart{\'\i}n} MC, {Gonz{\'a}lez} H, {Maestre Garc{\'\i}a} JL, {Marchant} J, {Ordonez-Etxeberria} I, {Martorell} P, {Salamero} J, {Organero} F, {Ana} L, {Fonseca} F, {Peris} V, {Brevia} O, {Selva} A, {Perello} C, {Cabedo} V, {Gon{\c{c}}alves} R, {Ferreira} M, {Marques Dias} F, {Daassou} A, {Barkaoui} K, {Benkhaldoun} Z, {Guennoun} M, {Chouqar} J, {Jehin} E, {Rinner} C, {Lloyd} J, {El Moutamid} M, {Lamarche} C, {Pollock} JT, {Caton} DB,
  {Kouprianov} V, {Timerson} BW, {Blanchard} G, {Payet} B, {Peyrot} A, {Teng-Chuen-Yu} JP, {Fran{\c{c}}oise} J, {Mondon} B, {Payet} T, {Boissel} C, {Castets} M, {Hubbard} WB, {Hill} R, {Reitsema} HJ, {Mousis} O, {Ball} L, {Neilsen} G, {Hutcheon} S, {Lay} K, {Anderson} P, {Moy} M, {Jonsen} M, {Pink} I, {Walters} R, {Downs} B. 2022  {Constraints on the structure and seasonal variations of Triton's atmosphere from the 5 October 2017 stellar occultation and previous observations}. {\em \aap} \textbf{659}, A136.
(\href{http://dx.doi.org/10.1051/0004-6361/202141443}{10.1051/0004-6361/202141443})

\bibitem{TMT2022}
{Sallum} S, {Millar-Blanchaer} MA, {Batalha} N, {Wang} J, {Martinez} R, {Fitzgerald} MP, {Skemer} A, {Jensen-Clem} R, {Mazin} BA, {Chun} M, {Guyon} O, {Hinz} P, {Males} J, {Max} C. 2022  {The Planetary Systems Imager for TMT: driving science cases and top level requirements}. In {Evans} CJ, {Bryant} JJ, {Motohara} K, editors, {\em Ground-based and Airborne Instrumentation for Astronomy IX} vol. 12184{\em Society of Photo-Optical Instrumentation Engineers (SPIE) Conference Series} p. 1218446. International Society for Optics and Photonics SPIE.
(\href{http://dx.doi.org/10.1117/12.2630423}{10.1117/12.2630423})

\bibitem{Sicardy19}
{Sicardy} B, {Leiva} R, {Renner} S, {Roques} F, {El Moutamid} M, {Santos-Sanz} P, {Desmars} J. 2019  {Ring dynamics around non-axisymmetric bodies with application to Chariklo and Haumea}. {\em Nature Astronomy} \textbf{3}, 146--153.
(\href{http://dx.doi.org/10.1038/s41550-018-0616-8}{10.1038/s41550-018-0616-8})

\end{thebibliography}

\end{document}